\documentclass[aps,reprint,superscriptaddress,nofootinbib,showpacs]{revtex4}
\usepackage{amsmath,latexsym,amssymb,hyperref,graphicx}
\usepackage{slashed}

\hypersetup{colorlinks,citecolor= nicegreen,linkcolor= niceblue}
\usepackage{color}
\definecolor{nicered}{rgb}{0.7,0.1,0.5}
\definecolor{nicegreen}{rgb}{0.1,0.5,0.1}
\definecolor{niceblue}{rgb}{0.1,0.1,0.8}

\begin{document}

\title{Conformal Isometries and Curvature Collineations of an Impulsive Plane Wave:\\ a distributional approach}
 
 \author{Juan Calles}\email{juan.calles.h@mail.pucv.cl}
\affiliation{Instituto de F\'isica, Pontificia Universidad Cat\'olica de Valpara\'iso, Casilla 4950, Valpara\'iso, Chile}
\author{Nelson Pantoja}\email{pantoja@ula.ve}
\affiliation{Centro de F\'isica Fundamental, Universidad de Los Andes, M\'erida 5101, Venezuela}


\begin{abstract}
\noindent

By extending the notion of Lie derivative to distribution-valued tensor fields of order $m$, Lie derivatives with respect to $C^k$ vector fields, $k\geqslant m+1$, can be shown to be well defined. Geometric symmetries, definable in terms of these Lie derivatives, can then be considered. In particular, for spacetimes of low regularity, geometric symmetries generated by vector fields of regularity as low as $C^1$ are definable. We find that the conformal algebra of the impulsive plane wave spacetime (with $+$ polarization) described by the continuous Baldwin-Jeffery-Rosen form of the metric has the maximal dimension seven. The curvature of this metric is well defined as a distribution and since this spacetime is Ricci-flat, we show that all the vector fields in the conformal algebra are special curvature collineations. We find also the general form of these last ones. Finally, we find the conformal Killing vector fields of this spacetime when using the distributional Brinkmann form of the metric.
        
\end{abstract}


\maketitle

\section{Introduction}

A spacetime in general relativity is usually referred as a pseudo-Riemannian manifold 
$(\mathcal{M},\mathbf{g})$ where $\mathbf{g}$ is a smooth Lorentz metric. However, 
spacetimes of low regularity, revealing itself through a lack of smoothness of the metric 
$\mathbf{g}$ and describing situations in which there are curvature concentrations in 
lower dimensional regions, are also considered physically relevant. Well known examples 
include black holes, gravitational waves and shock waves. Indeed, for the computation 
of the curvature of a spacetime of low regularity, one must resort to mathematical tools  
\cite{israel,barrabes,geroch,grosser} not within the classical smooth Lorentzian geometry.
On the other hand, issues of low regularity concerning to geodesics, singularity theorems and 
causality theory have been the object of intense investigation (see \cite{Graf:2019wuk,Grant:2019rye,Garcia-Heveling:2021cel} and references therein).

It is well known that geometric symmetries definable in terms of Lie derivatives give rise 
to constants of motion which are useful for integrating the geodesic equations in a given 
spacetime $(\mathcal{M},\mathbf{g})$. Up to what extent can the classical concept of a 
geometric symmetry (in the smooth case) be transferred to a spacetime of low regularity? 
Geometric symmetries of spacetimes of low regularity 
has been explored only in a few specially interesting cases. Killing symmetries of distributional 
metrics have been considered in Ref.\cite{Aichelburg:1994su}, 
where it is shown that the Killing fields of the Schwarzschild metric are 
also Killing fields of its ultra-relativistic limit, the last one being an impulsive pp-wave 
\cite{Aichelburg:1970dh}. Based on the analysis of the adjoint 
orbits of normal-form-preserving diffeomorphisms, Killing symmetries of impulsive 
pp-waves have been analyzed in Ref.\cite{aichelburg95} and the existence of non-smooth Killing vectors 
put forward in Ref.\cite{aichelburg96}. Lie derivatives of distribution-valued tensor fields 
have been used in Ref.\cite{Pantoja:2003zr} to study the permanence and/or the rising 
of geometric symmetries in the thin wall limit of domain wall spacetimes 
\cite{Guerrero:2002ki,Melfo:2002wd}.

Motivated by the fact that impulsive pp-waves have been frequently used as test models 
in non-smooth Lorentzian geometry and expecting to get further insight about the nature 
of these spacetimes by a better understanding of their symmetries, in this work we will 
focus on conformal isometries and curvature collineations of impulsive plane waves. 

By definition, a pp-wave (plane fronted wave with parallel rays) metric admits a covariantly constant null Killing vector, namely $\boldsymbol{\partial}_V$. In four dimensions, a pp-wave metric can always be written in the Brinkmann form
\begin{equation}\label{pp_wave}
A(U,X,Y)\mathbf{d}U\otimes\mathbf{d}U -\mathbf{d}U\otimes\mathbf{d}V -\mathbf{d}V\otimes\mathbf{d}U +\mathbf{d}X\otimes\mathbf{d}X + \mathbf{d}Y\otimes\mathbf{d}Y,
\end{equation}
where $(X,Y)$ and $(U,V)$ are the transversal and null coordinates, respectively, and $A$ any smooth function. For $A$ a harmonic function on the wavefront, $(\partial_X^2 +\partial_Y^2)A=0$, the Ricci and Einstein curvatures are zero. 

A plane wave metric is a pp-wave metric (\ref{pp_wave}) with
\begin{equation}\label{plane_wave}
A(U,X,Y)=A(U)F(X,Y),
\end{equation}
$A(U)$ any smooth function and $F(X,Y)$ quadratic in the transversal coordinates $(X,Y)$. For $F(X,Y)$ given by
\begin{equation}\label{plus_polarization}
F(X,Y)=\left(X^2 -Y^2\right),
\end{equation}
we have a plane wave with the profile fixed to the $+$ polarization.
 
A particularly interesting type of a 4-dimensional plane wave spacetime is the spacetime $(\mathbb{R}^4, {^B\mathbf{g}})$ associated to an impulsive plane wave with $+$ polarization \cite{penrose,penrose_2,synge1972general}, whose metric $^B\mathbf{g}$ is given by
\begin{equation}\label{pp_brinkmann}
^B\mathbf{g}=\delta(U)(X^2 -Y^2)\mathbf{d}U\otimes\mathbf{d}U -\mathbf{d}U\otimes\mathbf{d}V -\mathbf{d}V\otimes\mathbf{d}U +\mathbf{d}X\otimes\mathbf{d}X + \mathbf{d}Y\otimes\mathbf{d}Y,
\end{equation}
where $\delta(U)$ is the Dirac measure with support on the surface $U=0$. Obviously, 
the metric (\ref{pp_brinkmann}) should be regarded as defined by its values on open sets 
in $\mathbb{R}^4$ as a distribution-valued $(0,2)$-tensor field. Indeed, it lies beyond the 
reach of  classical Lorentzian geometry.
 
The impulsive plane wave with $+$ polarization can also be described by the continuous 
Baldwin-Jeffery-Rosen form of the metric ${^R\mathbf{g}}$, given by
\begin{equation}\label{pprosen_2}
^R\mathbf{g}=-\mathbf{d}u \otimes\mathbf{d}v - \mathbf{d}v \otimes\mathbf{d}u + P^2(u)\,\mathbf{d}x \otimes\mathbf{d}x + Q^2(u)\,\mathbf{d}y \otimes\mathbf{d}y,\quad P(u)=\begin{cases} 1,\quad u<0\\ 1+u,u\geqslant 0\end{cases}\!\!\!\!,\quad Q(u)=\begin{cases}1,\quad u<0\\1-u,u\geqslant 0\end{cases}\!\!\!\!.
\end{equation}
Note that, apart from the coordinate singularity at $u=1$, ${^R\mathbf{g}}$ is smooth with 
respect to the variables $v,x,y$ but merely Lipschitz continuous with respect to $u$. Thus, it 
lies also beyond the reach of classical Lorentzian geometry but it is within the Geroch-Traschen 
class of metrics \cite{geroch} for which the curvature tensor is defined as a distribution.
 
It is well known that ${^B\mathbf{g}}$ and ${^R\mathbf{g}}$ may be related by the discontinuous change of coordinates \cite{synge1972general}
\begin{eqnarray}\label{trans}
U & =& u, \nonumber\\
V &=& v + \frac{1}{2}\Theta^+_{u} \left(x^2-y^2\right)+\frac{1}{2} u\Theta^+_{u} \left(x^2 +y^2\right),\nonumber\\
X &=& \left(1 + u \Theta^+_{u}\right)x ,\nonumber\\
Y &=& \left(1 - u \Theta^+_{u}\right)y,
\end{eqnarray}
where $\Theta^+_{u}$ is the Heaviside distribution with support on $u>0$. However, the discontinuous transformation (\ref{trans}) alters the differentiable structure of the manifold and the equivalence of $^B\mathbf{g}$ (\ref{pp_brinkmann}) and $^R\mathbf{g}$ (\ref{pprosen_2}) is by far not obvious. In the meantime, a lot of work have been devoted to stablish their suspected \textit{physical} equivalence. In particular, the geometry of these impulsive plane waves has been explored via the analysis of the geodesic equations using both the distributional (\ref{pp_brinkmann}) and the continuous form (\ref{pprosen_2}) of the metric. Within the context of nonlinear distributional geometry \cite{grosser}, the discontinuous transformation (\ref{trans}) have been shown to be the \textit{distributional shadow} of a generalized diffeomorphism that relates their null geodesics \cite{Kunzinger:1998nz,Kunzinger:1998xw} (see also Ref.\cite{Erlacher:2010ts}). We shall be interested in the symmetries of this impulsive plane wave and, despite the above physical equivalence, we shall maintain the superscripts in ${^B\mathbf{g}}$ and ${^R\mathbf{g}}$ for ease of the discussion. 
 
This paper is organized as follows. In section \ref{math_framework}, after an overview on the subject of distribution-valued tensor fields, the definition of Lie derivative of a distribution-valued tensor field of order m along a $C^{k}$ vector field, $k\geqslant m+1$, is given. 
In terms of this derivative, the Killing equation and its conformal generalization can be extended to singular spacetimes having tensor metrics of low differentiability, say $C^0$ or even distribution-valued metric tensor fields. Section \ref{CI_section} then contains as an application the analysis of Killing symmetries of the impulsive plane wave spacetime with metric ${^R\mathbf{g}}$, which were obtained in Ref.\cite{aichelburg96} by other means, and also obtain and discuss their conformal symmetries. The metric ${^R\mathbf{g}}$ (\ref{pprosen_2}) is within the maximal class of spacetime metrics allowing for a distributional curvature \cite{geroch} and since this spacetime is Ricci-flat, in section \ref{CC_section} we show that all the vector fields in the conformal algebra are special curvature collineations. Furthermore, we find the general form of a curvature collineation for this spacetime. In section \ref{brinkmann_form}, as a further application, the conformal isometries of the distributional Brinkmann form (\ref{pp_brinkmann}) are determined. The last section is devoted to summarize the results.

\section{Mathematical Framework}\label{math_framework}

\subsection{Distribution-valued tensor fields}\label{t_distributions}

For completeness, in this subsection we recall some basic definitions and properties of 
distribution-valued tensor fields used throughout this paper. We follow the approach of 
Refs.\cite{lichnerowicz,choquet} which resort to the use of a certain additional structure to maintain explicitly the covariance. 
 
Let $\mathcal{M}$ be a $C^\infty$ $n$-dimensional oriented paracompact manifold. Given a locally finite atlas on $\mathcal{M}$, we endow the space of $C^\infty$  
$(p,q)$-tensor fields with compact support on $\mathcal{M}$ with an inductive limit of Fr\'echet topologies, i.e. with a Schwartz topology. Let $\mathcal{D}(\mathcal{M}, \otimes^p_q)$ be the space of such tensor fields with a Schwartz topology. The space of distribution-valued $(q,p)$-tensors on $\mathcal{M}$, $\mathcal{D}^{'}(\mathcal{M},\otimes^q_p)$, is the dual of
$\mathcal{D}(\mathcal{M},\otimes^p_q)$. Thus, a distribution-valued $(q,p)$-tensor 
$\mathbf{T}$ on $\mathcal{M}$ is a 
continuous linear functional on the test space $\mathcal{D}(\mathcal{M},\otimes^p_q)$: if $\mathbf{T}\in \mathcal{D}^{'}(\mathcal{M},\otimes^q_p)$ and $\mathbf{U}\in\mathcal{D}(\mathcal{M},\otimes^p_q)$, 
\begin{equation}\label{linear_mapping}
\mathbf{T}: \mathcal{D}(\mathcal{M},\otimes^p_q)\rightarrow\mathbb{R}\,\,\,
\text{by}\, \,\,\mathbf{U}\mapsto \langle \mathbf{T},\mathbf{U}\rangle.
\end{equation}

Let us endow $\mathcal{M}$ with a $C^{\infty}$ riemannian metric $\boldsymbol{g}$ 
and let ${^{\boldsymbol{g}}\nabla}$ be the torsion free metric compatible Levi-Civita 
covariant derivative associated with $\boldsymbol{g}$.  We denote by 
${^{\boldsymbol{g}}}\nabla\mathbf{U}$ the derivative in $\boldsymbol{g}$ of the tensor 
field $\mathbf{U}$ and by $\|\mathbf{U}(x)\|$ the pointwise norm (with respect to 
$\boldsymbol{g}$) of $\mathbf{U}$,
\begin{equation}
\|\mathbf{U}(x)\|=|U^{\alpha_1\ldots \alpha_p}_{\qquad \beta_1\ldots \beta_q}(x)\,U_{\alpha_1\ldots \alpha_p}^{\qquad \beta_1\ldots \beta_q}(x)|^{1/2},
\end{equation}
where indices are raised or lowered using the metric $\boldsymbol{g}$.  

The linear mapping (\ref{linear_mapping}) is continuous if and only if for 
each $\mathbf{U}$ with support in the domain $(K,\phi)$ of an 
admissible chart, there exists a constant $C(K)$ and a non negative integer $m(K)$ such that 
\begin{equation}\label{continuous_linear_mapping}
|\langle \mathbf{T},\mathbf{U}\rangle|\leqslant C(K)\, p_{K,m(K)}(\mathbf{U}),\quad p_{K,m(K)}(\mathbf{U})\equiv \sum_{|l|\leqslant m(K)} \,\text{sup}_{x\in K}\| ({^{\boldsymbol{g}}\nabla}^{(l)}\mathbf{U})(x)\|,
\end{equation}
where $l$ is a non negative multi index. The topology induced by the family of seminorms 
$p_{K,m(K)}(\mathbf{U})$ is independent of the choice of the metric $\boldsymbol{g}$. Hence, 
the spaces $\mathcal{D}(\mathcal{M},\otimes^p_q)$ and 
$\mathcal{D}^{'}(\mathcal{M},\otimes^q_p)$ will be independent of the choice of 
$\boldsymbol{g}$. 

The characterization (\ref{continuous_linear_mapping}) leads to the concept of the order of 
a distribution. We define the local order of $\mathbf{T}$ on $K$ as the minimum of all natural numbers $m(K)$ for which (\ref{continuous_linear_mapping}) $m$ holds. The order 
of $\mathbf{T}$ is the supremum over all local orders. 

The space of distribution-valued $(q,p)$-tensor fields of order $m$ is defined as follows.  
Let $\mathcal{D}^{m}_{K}(\mathcal{M},\otimes^p_q)$ be the space 
of $C_c^{m}$ $(p,q)$-tensor fields on $\mathcal{M}$ with compact support $K$ endowed 
with the topology induced by the family of seminorms $p_{K,m(K)}(\mathbf{U})$ defined in (\ref{continuous_linear_mapping}).
Let $\mathcal{D}^{m}(\mathcal{M},\otimes^p_q)$ be the space of $C_c^{m}$
$(p,q)$-tensor fields on $\mathcal{M}$ with compact support endowed
with the ``locally convex inductive limit''\footnote{For the definition and the topology of $\mathcal{D}^{m}$ see Ref.\cite{schwartz}} of the topologies 
$\mathcal{D}^{m}_{K}(\mathcal{M},\otimes^p_q)$. The space of distribution-valued $(q,p)$-tensor fields of order $m$ on $\mathcal{M}$, $\mathcal{D}^{'m}(\mathcal{M},\otimes^q_p)$, is the dual of $\mathcal{D}^{m}(\mathcal{M},\otimes^p_q)$. We have the following inclusions
\[
\mathcal{D}(\mathcal{M},\otimes^p_q)\hookrightarrow\mathcal{D}^{k}(\mathcal{M},\otimes^p_q)\hookrightarrow\mathcal{D}^{m}(\mathcal{M},\otimes^p_q),\qquad m\leqslant k,
\]
\[
\mathcal{D}^{'m}(\mathcal{M},\otimes^q_p)\hookrightarrow\mathcal{D}^{'k}(\mathcal{M},\otimes^q_p)\hookrightarrow\mathcal{D}^{'}(\mathcal{M},\otimes^q_p),\qquad m\leqslant k.
\]
A distribution-valued tensor field $\mathbf{T}\in \mathcal{D}^{'}(\mathcal{M},\otimes^q_p)$ is of order $m$ if and only if it can be extended to a linear continuous form on $\mathcal{D}^{m}(\mathcal{M},\otimes^p_q)$, that is if and only if $\mathbf{T}\in\mathcal{D}^{'m}(\mathcal{M},\otimes^q_p)$.

Just as for scalar distributions where the product of a scalar distribution of order $m$ by a $C^m$ function is defined, the tensor product of a distribution-valued tensor field of order $m$ and a $C^m$ tensor field is  defined. Tensor products of distribution-valued tensor fields are not definable in general.

The identification of a locally integrable $(q,p)$-tensor field $\mathbf{T}$ with a 
distribution-valued tensor is obtained \textit{via}
\begin{equation}
\langle{\bf T},{\bf U}\rangle\equiv \int_{\cal M} ({\bf T}|{\bf
U})\,\boldsymbol{\omega}_{\boldsymbol{g}},\qquad \forall \,{\bf{U}} \in \mathcal{D}(\mathcal{M},\otimes^p_q), \label{Tdist}
\end{equation}
where $({\bf T}|{\bf U})$ denotes the scalar product of $\mathbf{T}$ and $\mathbf{U}$, and $\boldsymbol{\omega}_{\boldsymbol{g}}$ is
the volume element of $\boldsymbol{g}$. Since (\ref{Tdist}) is
the integral of an $n$-form with compact support, in the domain of a chart 
$(\mathcal{M},\phi)$ we have
\begin{equation}\label{Tdist2}
\langle{\bf T},{\bf U}\rangle\equiv \int_{\phi({\cal M})} T_{\qquad \beta_1\ldots
\beta_p}^{\alpha_1\dots \alpha_q}U^{\beta_1\ldots \beta_p}_{\qquad \alpha_1\dots \alpha_q} |\det g|^{\frac{1}{2}} dx^1\ldots dx^n .
\end{equation}
Let $\mathbf{U}\in \mathcal{D}(\mathcal{M},\otimes^p_q)$ with support in $K$. From (\ref{Tdist2}), 
we have
\begin{equation}\label{continuous_linear_mapping_2}
\left| \langle{\bf T},{\bf U}\rangle \right| \leqslant C (K) \,\text{sup}_{x\in K} \|\mathbf{U}\|,\qquad
C(K)= \int_{\phi(K)} \|\mathbf{T} \|\,| \det g|^{\frac{1}{2}} dx^1\ldots dx^n.
\end{equation}
This mapping is linear and continuous for every compact subset $K\in\mathcal{M}$. From (\ref{continuous_linear_mapping}) and (\ref{continuous_linear_mapping_2}), it follows that a locally integrable tensor field $\mathbf{T}$ on $\mathcal{M}$ defines a distribution-valued tensor on $\mathcal{M}$ of order 0. 

For $\mathbf{T}$ an arbitrary $(q,p)$-tensor distribution, the covariant derivative in the metric $\boldsymbol{g}$ is the $(q,p+1)$-tensor ${^{\boldsymbol{g}}\boldsymbol{\nabla}} \mathbf{T}$ defined by
\begin{equation}\label{cov_deriv}
\langle {^{\boldsymbol{g}}\boldsymbol{\nabla}} \mathbf{T},\mathbf{U}\rangle\equiv -\langle \mathbf{T},\boldsymbol{g}\cdot{^{\boldsymbol{g}}\boldsymbol{\nabla}}\mathbf{U}\rangle,\qquad \forall\, \mathbf{U}\in\mathcal{D}(\mathcal{M},\otimes^{p+1}_q),
\end{equation}
where $\boldsymbol{g}\cdot{^{\boldsymbol{g}}\boldsymbol{\nabla}} \mathbf{U}$ is the contracted product of $\boldsymbol{g}$ with ${^{\boldsymbol{g}}\boldsymbol{\nabla}} \mathbf{U}$, that is, in a coordinate chart
\[
(\boldsymbol{g}\cdot{^{\boldsymbol{g}}\boldsymbol{\nabla}} \mathbf{U})^{\beta_1\ldots \beta_p}={^{\boldsymbol{g}}\nabla}_\alpha U^{\alpha\beta_1\ldots \beta_p}_{\qquad \,\,\,\,\alpha_1\dots \alpha_q}.
\] 
If $\mathbf{T}$ is a distribution-valued tensor field of order $m$ on a compact set then ${^{\boldsymbol{g}}\boldsymbol{\nabla}}\mathbf{T}$ is a distribution-valued tensor field of order $\leqslant m+1$ on the same compact set. Note that the components of ${^{\boldsymbol{g}}\boldsymbol{\nabla}} \mathbf{T}$ in a coordinate chart are given by the usual formulas.

Let $\mathit{\Sigma}$ be a $(n-1)$-dimensional submanifold of ${\cal M}$. Let us suppose that $\mathit{\Sigma}$ is defined by the equation $\Phi=0$, with $\Phi: {\cal M}\rightarrow \mathbb{R}$ a smooth function without critical points. We suppose that $\mathcal{M}$ is divided in disjoint open sets $\mathcal{M}_+\equiv \{x\in \mathcal{M}\,|\, \Phi(x)>0\}$, $\mathcal{M}_-\equiv \{x\in \mathcal{M}\,|\, \Phi(x)<0\}$. We fix an orientation by setting $\Sigma\equiv \partial\Omega_-$.
The Dirac measure $\delta_\Sigma$ is the distribution of order 0 with support $\Sigma$ defined by
\[
\langle \delta_\Sigma,\varphi\rangle\equiv \int_\Sigma \varphi \,\boldsymbol{\sigma},\qquad \forall \varphi\in\mathcal{D}(\mathcal{M}),
\]
where $\boldsymbol{\sigma}$ is the $(n-1)$-form relative to $\Sigma$, called the Leray form of $\Sigma$, defined by
\[
\boldsymbol{\omega}_{\boldsymbol{g}}\equiv \mathbf{d}\Phi\wedge \boldsymbol{\sigma}.
\] 
Since $\Phi$ has non-vanishing gradient, we can choose charts on $\mathcal{M}$ such that $\Phi(x)\equiv x^1$ and 
\[
\sigma\equiv \rho \,\mathbf{d}x^2\wedge\cdots\wedge\mathbf{d}x^n.
\]

\subsection{Distributional Lie derivatives}\label{lie_distributions}

Given appropriate definitions of distribution-valued tensor fields and their covariant derivatives, we now address the analogous question for the Lie derivative of a distribution-valued tensor. Before giving the precise definitions, we will motivate these by looking at the Lie derivatives of $C^1$ tensor fields as distribution-valued tensor fields. To begin with, 
as in the previous subsection \ref{t_distributions}, suppose that the manifold $\mathcal{M}$ is endowed with a $C^\infty$ metric $\boldsymbol{g}$, with $\boldsymbol{\omega}_{\boldsymbol{g}}$ its volume element  and ${^{\boldsymbol{g}}\nabla}$ its Levi-Civita connection. Let $\mathbf{T}$ be a $C^1$ $(q,p)$-tensor field on $\mathcal{M}$, let 
$\mathbf{V}$ be a smooth vector field on
$\mathcal{M}$. The Lie derivative of $\mathbf{T}$ with respect to $\mathbf{V}$ is the
$(q,p)$-tensor field $\mathcal{L}_{_\mathbf{V}} \mathbf{T}$ such
that,
\begin{eqnarray}
\langle\mathcal{L}_{_\mathbf{V}} \mathbf{T},{\bf U}\rangle=&& \int_{\mathcal{M}} (\mathcal{L}_{_{\mathbf{V}}}\mathbf{T}|\mathbf{U})\,\boldsymbol{\omega}_{\boldsymbol{g}}\nonumber\\=&&\int_{\mathcal{M}}\mathcal{L}_{_{\mathbf{V}}}((\mathbf{T}|\mathbf{U})\,\boldsymbol{\omega}_{\boldsymbol{g}})-\int_\mathcal{M}
\left(({\bf T} |\mathcal{L}_{_\mathbf{V}}{\bf U})
\boldsymbol{\omega}_{\boldsymbol{g}} + (\mathbf{T}|\mathbf{U})
\mathcal{L}_{_\mathbf{V}}\boldsymbol{\omega}_{\boldsymbol{g}}\right),\quad\forall \,\mathbf{U}\in \mathcal{D}(\mathcal{M},\otimes^p_q). \label{liederivsmooth}
\end{eqnarray}
Now, since $({\bf T}\cdot{\bf U})\, \boldsymbol{\omega}_{\boldsymbol{g}}$ is
an $n$-form with compact support, we have
\begin{equation}
\int_\mathcal{M} \mathcal{L}_{_\mathbf{V}}(({\bf T}|{\bf U})\,
\boldsymbol{\omega}_{\boldsymbol{g}})= \int_\mathcal{M}
{\mathbf{di}}_{_\mathbf{V}}(({\bf T}|{\bf U})\,
\boldsymbol{\omega}_{\boldsymbol{g}})= \int_{\partial
\mathcal{M}}{\mathbf{i}}_{_\mathbf{V}}(({\bf T}|{\bf U})\,
\boldsymbol{\omega}_{\boldsymbol{g}})= 0,
\end{equation}
where ${\mathbf{i}}_{_\mathbf{V}}$ denotes interior product, in
the last step we have used Stokes' theorem and the surface term
vanishes because $\mathbf{U}$ has compact support. On the other hand,
\begin{equation}\label{liederiv_vol}
\mathcal{L}_{_\mathbf{V}}\boldsymbol{\omega}_{\boldsymbol{g}}=\frac{1}{2}\text{Tr}\,\{\mathcal{L}_{_{\mathbf{V}}}\boldsymbol{g}\}\boldsymbol{\omega}_{\boldsymbol{g}},
\end{equation}
where Tr stands for ``trace of" relative to $\boldsymbol{g}$, i.e.
\begin{equation}\label{trace}
\text{Tr}\,\{{\mathcal{L}_{_{\mathbf{V}}}\boldsymbol{g}\}}=(\boldsymbol{g}^{-1}|\mathcal{L}_{_{\mathbf{V}}}\boldsymbol{g}).
\end{equation}

Thus, given on $\mathcal{M}$ a $C^\infty$ volume element $\boldsymbol{\omega}_{\boldsymbol{g}}$, the Lie derivative $\mathcal{L}_{_\mathbf{V}}\mathbf{T}$ of an ordinary $C^1$ $(q,p)$-tensor $\mathbf{T}$ along a $C^{\infty}$ vector field $\mathbf{V}$ satisfies
\begin{equation}
\langle\mathcal{L}_{_\mathbf{V}} \mathbf{T},{\bf U}\rangle= -\int_\mathcal{M}
\left({\bf T} \left| \left(\mathcal{L}_{_\mathbf{V}} + \frac{1}{2}(\boldsymbol{g}^{-1}|\mathcal{L}_{_{\mathbf{V}}}\boldsymbol{g})\right)
\mathbf{U}\right.\right)\boldsymbol{\omega}_{\boldsymbol{g}}.\label{rhs}
\end{equation}
 This leads us to adopt the following definition:

For $\mathbf{T}$ an arbitrary $(q,p)$-tensor distribution and $\mathbf{V}$ a
$C^{\infty}$ vector field on $(\mathcal{M},\boldsymbol{g})$, the Lie derivative $\mathcal{L}_{_\mathbf{V}}\mathbf{T}$ is the $(q,p)$-tensor distribution given by
\begin{equation}
\langle\mathcal{L}_{_\mathbf{V}} \mathbf{T},{\bf U}\rangle\equiv -
\left\langle\mathbf{T},\left(\mathcal{L}_{_\mathbf{V}} +
\frac{1}{2}(\boldsymbol{g}^{-1}|\mathcal{L}_{_{\mathbf{V}}}\boldsymbol{g}) \right)\mathbf{U}\right\rangle,\qquad \forall \,\mathbf{U}\in \mathcal{D}(\mathcal{M},\otimes^p_q).\label{liederiv_a}
\end{equation}

Note that if $\mathbf{U}\in \mathcal{D}(\mathcal{M},\otimes^p_q)$, for $\mathbf{V}$ a $C^{\infty}$ vector field on $\mathcal{M}$ we have  
$\left(\mathcal{L}_{_\mathbf{V}} +(1/2)(\boldsymbol{g}^{-1}|\mathcal{L}_{_{\mathbf{V}}}\boldsymbol{g})\right)\mathbf{U}\in \mathcal{D}(\mathcal{M},\otimes^p_q)$ and the mapping
\[
\mathbf{U}\mapsto - \left\langle\mathbf{T},\left(\mathcal{L}_{_\mathbf{V}} +
({1}/{2})(\boldsymbol{g}^{-1}|\mathcal{L}_{_{\mathbf{V}}}\boldsymbol{g}) \right)\mathbf{U}\right\rangle
\]
is linear and continuous on $\mathcal{D}(\mathcal{M},\otimes^p_q)$. Indeed, as follows directly from the derivation of (\ref{rhs}), the distributional Lie derivative (\ref{liederiv_a}) reduce to the classical one when $\mathbf{T}$ is a $C^1$ tensor field on $\mathcal{M}$. 

Let $\nabla$ be any smooth torsion free covariant derivative on $\mathcal{M}$. We have
\begin{eqnarray}\label{trace_2}
(\boldsymbol{g}^{-1}|\mathcal{L}_{_{\mathbf{V}}}\boldsymbol{g})=(\boldsymbol{g}^{-1})^{\alpha\beta}(\mathcal{L}_{_{\mathbf{V}}}\boldsymbol{g})_{\alpha\beta}=&&\left[(g^{\alpha\beta}V^\gamma{\nabla}_\gamma g_{\alpha\beta} +2{\nabla}_\alpha V^\alpha\right]\nonumber\\=&&2\,({^{\boldsymbol{g}}\nabla}\cdot\mathbf{V}),
\end{eqnarray}
where the second line of (\ref{trace_2}) holds when $\nabla$ is the Levi-Civita covariant derivative  
${^{\boldsymbol{g}}\nabla}$ associated to $\boldsymbol{g}$ and we recover the definition of  distributional Lie derivative of Ref.\cite{Pantoja:2003zr}.

We are now going to consider the case where $\mathbf{V}$ is a non-smooth vector 
field on $\mathcal{M}$ for which, in general,  $\mathcal{L}_{_\mathbf{V}} \mathbf{U}$ 
and $(\boldsymbol{g}^{-1}|\mathcal{L}_{_{\mathbf{V}}}\boldsymbol{g})\mathbf{U}$ are not in $\mathcal{D}_{(p,q)}(\mathcal{M})$. With this purpose, turn our attention again to the Lie derivative of a $C^1$ tensor field $\mathbf{T}$ with respect to a smooth vector field $\mathbf{V}$. In a natural basis, for a $C^1$ $(q,p)$-tensor $\mathbf{T}$ of the form
\[
\mathbf{T}=T_{\qquad \beta_1\ldots \beta_p}^{\alpha_1\dots \alpha_q} \boldsymbol{\partial}_{x^{\alpha_1}}\otimes\cdots\otimes\boldsymbol{\partial}_{x^{\alpha_q}}\otimes\mathbf{d}x^{\beta_1}\otimes\cdots\otimes\mathbf{d}x^{\beta_p},
\]
the Lie derivative $\mathcal{L}_{_\mathbf{V}} \mathbf{T}$ is given by 
\begin{eqnarray}\label{liederiv_smooth}
\mathcal{L}_{_\mathbf{V}} \mathbf{T}&=&\left(V^\gamma\partial_\gamma T_{\qquad \beta_1\ldots \beta_p}^{\alpha_1\dots \alpha_q} + T_{\qquad \gamma \beta_2\ldots \beta_p}^{\alpha_1\dots \alpha_q}\partial_{\beta_1}V^\gamma +\cdots + T_{\qquad \beta_1\ldots \beta_{p-1}\gamma}^{\alpha_1\dots \alpha_q}\partial_{\beta_p}V^\gamma\right.\nonumber\\
&&\left.\quad - T_{\qquad \,\,\,\,\beta_1\ldots \beta_p}^{\gamma \alpha_2\dots \alpha_q}\partial_{\gamma}V^{\alpha_1} - \cdots - T_{\qquad \quad\,\,\,\,\,\beta_1\ldots \beta_p}^{\alpha_1\dots \alpha_{q-1}\gamma}\partial_{\gamma}V^{\alpha_q} \right) \boldsymbol{\partial}_{x^{\alpha_1}}\otimes\cdots\otimes\boldsymbol{\partial}_{x^{\alpha_q}}\otimes\mathbf{d}x^{\beta_1}\otimes\cdots\otimes\mathbf{d}x^{\beta_p},
\end{eqnarray}
where all the partial derivatives $\partial_\gamma T_{\qquad \beta_1\ldots \beta_p}^{\alpha_1\dots \alpha_q}$, $\partial_{\beta_p}V^\gamma$, $\dots$ , can be replaced by the components of the torsion free Levi-Civita covariant derivatives ${^{\boldsymbol{g}}\nabla}_\gamma T_{\qquad \beta_1\ldots \beta_p}^{\alpha_1\dots \alpha_q}$, $\ldots$ , associated to an arbitrary $C^\infty$ riemannian metric $\boldsymbol{g}$, in order to make explicit the fact that the result is indeed a tensor field.\footnote{Recall that classically, among the main properties of the Lie derivative, it transforms tensors of a given type into tensors of the same given type and it is connection-independent.} Next, let $T_{\qquad \beta_1\ldots \beta_p}^{\alpha_1\dots \alpha_q}$ and their derivatives ${^{\boldsymbol{g}}\nabla}_\gamma T_{\qquad \beta_1\ldots \beta_p}^{\alpha_1\dots \alpha_q}$ be the components of the distribution-valued tensor fields of order $0$ and $\leqslant1$ defined in (\ref{Tdist}) and (\ref{cov_deriv}), respectively. Indeed, the resulting expression for $\mathcal{L}_{_\mathbf{V}} \mathbf{T}$ can be interpreted as a distribution whenever each term can be interpreted as a distribution. This requires that both tensor products $\mathbf{V}\otimes{^{\boldsymbol{g}}\boldsymbol{\nabla}}\mathbf{T}$ and ${^{\boldsymbol{g}}\boldsymbol{\nabla}}\mathbf{V}\otimes\mathbf{T}$ be distribution-valued tensor fields, where only the factors ${^{\boldsymbol{g}}\boldsymbol{\nabla}}\mathbf{T}$ and $\mathbf{T}$ can be distributions. Thus, if we want to define the distributional Lie derivative of a distribution-valued tensor field $\mathbf{T}$ with respect to a non-smooth vector field $\mathbf{V}$ we would multiply a distribution with a non-smooth function which is problematic. However, from the fact that the tensor product of a distribution-valued tensor field of order $m$ and a $C^m$ tensor field is defined, we find that in this case the vector field $\mathbf{V}$ should be of at least $C^1$ regularity on $\mathcal{M}$. It follows that $\mathcal{L}_{_\mathbf{V}} \mathbf{T}$ is a distribution-valued tensor of order $\leqslant1$ and the identification (\ref{rhs}) holds for $\mathbf{U}\in \mathcal{D}^{1}(\mathcal{M},\otimes^p_q)$ since $\mathcal{L}_{_\mathbf{V}} \mathbf{U}+(1/2)(\boldsymbol{g}^{-1}|\mathcal{L}_{_{\mathbf{V}}}\boldsymbol{g})\mathbf{U}\in \mathcal{D}^{0}(\mathcal{M},\otimes^p_q)$.
 
Thus, in order to define the Lie derivative of a distribution-valued tensor field with respect to a non-smooth vector field, the above lead us to restrict ourselves to subspaces of $\mathcal{D}^{'}(\mathcal{M},\otimes^q_p)$ where the following definition makes sense: 

Let $\mathbf{T}$ be a distribution-valued $(q,p)$-tensor field on $(\mathcal{M},\boldsymbol{g})$ of order $m$. Then for $\mathbf{V}$ a $C^{k}$ vector field on $(\mathcal{M},\boldsymbol{g})$, $k\geqslant m+1$, the Lie derivative $\mathcal{L}_{_\mathbf{V}}\mathbf{T}$ is the distribution-valued $(q,p)$-tensor of order $\leqslant m+1$ defined by
\begin{equation}
\langle\mathcal{L}_{_\mathbf{V}} \mathbf{T},{\bf U}\rangle\equiv -
\left\langle\mathbf{T},\left(\mathcal{L}_{_\mathbf{V}} +
\frac{1}{2}(\boldsymbol{g}^{-1}|\mathcal{L}_{_{\mathbf{V}}}\boldsymbol{g})\right) \mathbf{U}\right\rangle,\qquad \forall \,\mathbf{U}\in \mathcal{D}(\mathcal{M},\otimes^p_q).\label{liederiv_b}
\end{equation}

We can check that a distribution-valued $(q,p)$-tensor of order $\leqslant m+1$ is really 
defined in this way. Being $\mathbf{T}\in \mathcal{D}^{'m}(\mathcal{M},\otimes^q_p)$, for $\mathbf{V} \in C^k(\mathcal{M})$, $k\geqslant m+1$, the 
RHS of (\ref{liederiv_b}) is defined  if $\mathbf{U}\in \mathcal{D}^{m+1}(\mathcal{M},\otimes^p_q)$. Obviously, for distribution-valued tensor fields of order $m$, the case $\mathbf{V}$ of extremely low regularity, say $C^m$ or below, is beyond the reach of the distributional approach.
 
\section{Isometries and conformal isometries of the impulsive plane wave}\label{CI_section}

Let us recall the classical meaning of an isometry. For a spacetime
$(\mathcal{M},\mathbf{g})$ where $\mathbf{g}$ is
$C^{\infty}$ metric tensor, an isometry $\phi$ on $(\mathcal{M},\mathbf{g})$ is
defined to be a diffeomorphism $\phi:\mathcal{M}\rightarrow
\mathcal{M}$ for which
$\phi^*\mathbf{g}=\mathbf{g}$. The infinitesimal
generator of a one-parameter group $\phi_{\lambda}$ of local
isometries is the $C^{\infty}$ vector field $\mathbf{K}$ on
$\mathcal{M}$ that satisfies
\begin{equation}
\mathcal{L}_{_\mathbf{K}}\mathbf{g}=0,\label{KV}
\end{equation}
where $\mathcal{L}_{\mathbf{K}}$ is the Lie derivative along $\mathbf{K}$. The $C^{\infty}$ vector field $\mathbf{K}$ is called a Killing vector field on
$(\mathcal{M},\mathbf{g})$ relative to this group \cite{wald}. We shall denote by $\textsc{Isom}(\mathcal{M},\mathbf{g})$ the isometry group on $(\mathcal{M},\mathbf{g})$ and by $\mathfrak{Isom}(\mathcal{M},\mathbf{g})$ its algebra,
\begin{equation}
\mathfrak{Isom}(\mathcal{M},\mathbf{g})=\{\mathbf{K} : \mathcal{L}_{_\mathbf{K}}\mathbf{g}=0\}.
\end{equation}

Of greatest interest is the following generalization of the notion of Killing vector fields. A conformal isometry $\phi$ on $(\mathcal{M},\mathbf{g})$ is defined to be a diffeomorphism $\phi:\mathcal{M}\rightarrow \mathcal{M}$ for which there is a function $\Omega$ such that $\phi^*\mathbf{g}=\Omega^2\mathbf{g}$. An isometry is to be regarded as the special case $\Omega=1$ of a conformal transformation. A vector field $\mathbf{K}$ on $\mathcal{M}$ that satisfies 
\begin{equation}\label{CKV}
\mathcal{L}_{_{\mathbf{K}}}\mathbf{g}=2\psi\,\mathbf{g},
\end{equation}
is called a conformal Killing vector field of $\mathbf{g}$ if $\psi$ is not constant on $\mathcal{M}$. If $\psi$ is constant on $\mathcal{M}$ then $\mathbf{K}$ is called a homothetic Killing vector field and if $\psi=0$ then $\mathbf{K}$ is a Killing vector field. The set of all conformal Killing vector fields forms a finite dimensional algebra which will be denoted by $\mathfrak{Conf}(\mathcal{M},\mathbf{g})$. Conformal symmetries preserve the structure of the null cone by mapping null geodesics to null geodesics and the conformal Killing fields give rise to constants of motion for null geodesics \cite{wald}.

It is well known that for a smooth plane wave spacetime the maximum dimension of the algebra of conformal Killing vector fields is seven \cite{hall,Maartens:1991mj,Keane:2004dpc}. Five of them are isometries and there is one homothety. There may or may not exist a 7th transformation which may or may not be an isometry depending on special conditions. The isometry group of smooth generic plane gravitational waves has been identified as the Carroll group in (2+1) dimensions with broken rotations \cite{Duval:2017els}. Constants of motion for null geodesics associated with conformal isometries on smooth plane wave spacetimes have been discussed in Ref.\cite{Zhang:2019gdm}.

\subsection{Isometries of ${^R\mathbf{g}}$}    

We now revisit the isometries of the impulsive plane wave spacetime with metric ${^R\mathbf{g}}$ (\ref{pprosen_2}). The associated Killing vector fields were already obtained in Ref.\cite{aichelburg96} as the formal pullback under the discontinuous change of coordinates (\ref{trans}) of the Killing vector fields of ${^B\mathbf{g}}$ (\ref{pp_brinkmann}) obtained through a regularization scheme. Clearly, $^R\mathbf{g}$ depends neither on $v$ nor on the transverse coordinates $x,y$, so it is expected that there are at least three Killing vectors $\boldsymbol{\partial} _v, \boldsymbol{\partial} _x, \boldsymbol{\partial} _y$.
Indeed, all the Killing vector fields of the metric $^R\mathbf{g}$ can be found by solving the Killing equation 
\begin{equation}
\mathcal{L}_{_\mathbf{K}}{^R\mathbf{g}}=0.\label{KV_R}
\end{equation}
Admitting solutions of $C^1$ regularity across the surface $u=0$, we find

\begin{eqnarray}\label{ksuavespp3}
^R\mathbf{K}_1 =   \boldsymbol{\partial} _v,\qquad
&^R\mathbf{K}_2 =   \boldsymbol{\partial} _x,&\qquad
^R\mathbf{K}_3 =   \boldsymbol{\partial} _y,\quad\nonumber\\
^R\mathbf{K}_4 = \begin{cases} 
                                     x \boldsymbol{\partial} _v + u \boldsymbol{\partial}_x,\quad u<0\\
                                     x \boldsymbol{\partial} _v + (u/(1+u)) \boldsymbol{\partial}_x,\quad u\geqslant 0,
                                     \end{cases}\!\!\!\!&\quad&\!\!\!\!
^R\mathbf{K}_5 =  \begin{cases} 
                                     y \boldsymbol{\partial} _v + u \boldsymbol{\partial}_y,\quad u<0\\
                                     y \boldsymbol{\partial} _v + (u/(1-u)) \boldsymbol{\partial}_y,\quad u\geqslant 0.
                                     \end{cases}
\end{eqnarray}
The Killing vector fields in $\{^R\mathbf{K}_1,^R\mathbf{K}_2,^R\mathbf{K}_3\}$ are smooth solutions to (\ref{KV_R}) and satisfy 
\begin{equation}\label{trans_algebra}
[{^R\mathbf{K}}_i,{^R\mathbf{K}}_j]=0, \qquad i,j=1,2,3.
\end{equation}
The 3-dimensional abelian algebra spanned by these ones is a \textit{classical} translation algebra acting transitively on the null hypersufaces $\Sigma_u$ of constant $u$. 
On the other hand, the identification of the non smooth vector fields $^R\mathbf{K}_4$ and $^R\mathbf{K}_5$ as Killing vector fields of ${^R\mathbf{g}}$, assume a $C^1$ regularity across the surface $u=0$ of the solutions of (\ref{KV_R}).\footnote{Note that for the metric ${^R\mathbf{g}}$ (\ref{pprosen_2}), geodesics of $C^1$ regularity across the surface $u=0$ are definable \cite{Lecke:2013lja,Steinbauer:2013spa}.} Note also that ${^R\mathbf{K}}_5$ is unbounded at $u=1$, which can be traced back to the fact that ${^R\mathbf{g}}$ (\ref{pprosen_2}) exhibits a coordinate singularity at $u=1$.

To study the geometric properties of a manifold endowed with the non-smooth metric ${^R\mathbf{g}}$ (\ref{pprosen_2}), we will choose as a mathematical framework that of distribution-valued tensor fields. Let $\Omega$ be the open subset of $\mathbb{R}^4$ defined by 
\begin{equation}\label{Omega}
\Omega=\{(u,v,x,y)\in\mathbb{R}^4 | -u_0<u<u_0, (v,x,y)\in\mathbb{R}^3\},
\end{equation}
where $0<u_0<1$. For $(u,v,x,y)\in\Omega$, the metric $^R\mathbf{g}$ is smooth with respect to $(v,x,y)$ and is Lipschitz continuous in the variable $u$ at $u=0$. ${^R\mathbf{g}}$ (\ref{pprosen_2}) is therefore a locally integrable tensor field in $\Omega$ and defines a distribution-valued tensor field on $\Omega$ of order 0 \textit{via} (\ref{Tdist}). Note also that (\ref{pprosen_2}) is invertible on $\Omega$.  

We start by examining in detail the distributional meaning of (\ref{KV_R}) for the metric 
$^R\mathbf{g}$ along the vector fields (\ref{ksuavespp3}). Let $\boldsymbol{\eta}$ be the 
four-dimensional Minkowski metric tensor in light-cone coordinates
\begin{equation}
\boldsymbol{\eta}= -\mathbf{d}u\otimes\mathbf{d}v - \mathbf{d}v\otimes
\mathbf{d}u + \mathbf{d}x\otimes \mathbf{d}x + \mathbf{d}y\otimes
\mathbf{d}y.\label{minkowski}
\end{equation} 
Let $\mathbf{U}$ be a test $(2,0)$-tensor with compact support on $\Omega$. From the definition of Lie derivative (\ref{liederiv_a}), we have for the Killing vector $^R\mathbf{K}_2=\boldsymbol{\partial}_x$ and $\boldsymbol{g}=\boldsymbol{\eta}$ that $(\boldsymbol{\eta}^{-1}|\mathcal{L}_{_{^R\mathbf{K}_2}}\boldsymbol{\eta})=0$ and 
 \begin{eqnarray}
\langle\mathcal{L}_{_{^R\mathbf{K}_2}}{^R\mathbf{g}},{\bf U}\rangle\!&\equiv&\! -
\left\langle{^R\mathbf{g}},\,\mathcal{L}_{_{^R\mathbf{K}_2}} \mathbf{U} +
\frac{1}{2}(\boldsymbol{\eta}^{-1}|\mathcal{L}_{_{^R\mathbf{K}_2}}\boldsymbol{\eta}) \mathbf{U}\right\rangle\nonumber\\
\!&=&\!-\int_{-u_0}^0 du \int_{\mathbb{R}^3}dv\,dx\,dy \,\,\partial_x(-U^{uv} - U^{vu}+ U^{xx} +U^{yy}) \nonumber \\
\!&&\!-\int_{0}^{u_0} du\int_{\mathbb{R}^3}dv\,dx\,dy\,\,\partial_x(-U^{uv} -U^{vu}+ (1+u)^2U^{xx} +(1-u)^2U^{yy})= 0,
\end{eqnarray}
where we have used the fact that $\mathbf{U}$ is of compact support. It follows that $\mathcal{L}_{_{^R\mathbf{K}_2}}{^R\mathbf{g}}$ is the zero distribution on $\Omega$. Analogous calculations show that $\langle\mathcal{L}_{_{^R\mathbf{K}_1}}{^R\mathbf{g}},{\bf U}\rangle=0=\langle\mathcal{L}_{_{^R\mathbf{K}_3}}{^R\mathbf{g}},{\bf U}\rangle$, $\forall \mathbf{U}\in \mathcal{D}(\Omega,\otimes^2_0)$. 

Next, let us consider the non-smooth vector field $^R\mathbf{K}_4$, as given in (\ref{ksuavespp3}). To what extent can be considered $\mathcal{L}_{_{^R\mathbf{K}_4}}{^R\mathbf{g}}$ as a distribution? Define 
\begin{equation}\label{liederiv_zero_dist}
\langle\mathcal{L}_{_{^R\mathbf{K}_4}}{^R\mathbf{g}},{\bf U}\rangle\equiv -
\left\langle{^R\mathbf{g}},\,\mathcal{L}_{_{^R\mathbf{K}_4}} \mathbf{U} +
\frac{1}{2}(\boldsymbol{g}^{-1}|\mathcal{L}_{_{^R\mathbf{K}_4}}\boldsymbol{g}) \mathbf{U}\right\rangle,\qquad \mathbf{U} \in \mathcal{D}(\Omega,\otimes^2_0),
\end{equation}
where $\boldsymbol{g}$ is any smooth metric in $\Omega$. 
Indeed, this definition is meaningful if $\mathcal{L}_{_{^R\mathbf{K}_4}} \mathbf{U} +
\frac{1}{2}(\boldsymbol{g}^{-1}|\mathcal{L}_{_{^R\mathbf{K}_4}}\boldsymbol{g}) \mathbf{U}$ is in the space on which $^R\mathbf{g}$, as a distribution, is defined. Now, for $\mathbf{K}$ an arbitrary $C^1$ vector field on $\Omega$ and $\mathbf{U}\in \mathcal{D}^1(\Omega,\otimes^2_0)$ we have that $\mathcal{L}_{_{\mathbf{K}}} \mathbf{U}$ and $\frac{1}{2}(\boldsymbol{g}^{-1}|\mathcal{L}_{_{\mathbf{K}}}\boldsymbol{g}) \mathbf{U}$ are $C^0$ $(2,0)$-tensor fields with compact support on $\Omega$. Since the metric (\ref{pprosen_2}) is a distribution-valued tensor of order 0, $\langle\mathcal{L}_{_{^R\mathbf{K}_4}}{^R\mathbf{g}},{\bf U}\rangle$ is defined in the sense of (\ref{liederiv_b}).

For $\boldsymbol{g}=\boldsymbol{\eta}$, with $\boldsymbol{\eta}$ given by (\ref{minkowski}), we have $(\boldsymbol{\eta}^{-1}|\mathcal{L}_{_{^R\mathbf{K}_4}}\boldsymbol{\eta})=0$ and 
\begin{eqnarray}
\left\langle{^R\mathbf{g}},\,\mathcal{L}_{_{^R\mathbf{K}_4}} \mathbf{U}\right\rangle
\!&=&\!\int_{-u_0}^0 du \int_{\mathbb{R}^3}dv\,dx\,dy \,\,(x\partial_v +u\partial_x)(-U^{uv} - U^{vu}+ U^{xx} +U^{yy}) \nonumber \\
\!&&\!+\int_{0}^{u_0} du\int_{\mathbb{R}^3}dv\,dx\,dy\,\,(x\partial_v +\frac{u}{1+u}\partial_x)(-U^{uv} -U^{vu}+ (1+u)^2U^{xx} +(1-u)^2U^{yy})=0,
\end{eqnarray}
where we have used the fact that $\mathbf{U}$ is of compact support. In an analogous way it is found that $\mathcal{L}_{_{^R\mathbf{K}_5}}{^R\mathbf{g}}=0$ as a distribution on $\mathcal{D}(\Omega,\otimes^2_0)$.

Indeed, the results
\begin{equation}
\langle\mathcal{L}_{_{^R\mathbf{K}_i}}{^R\mathbf{g}},{\bf U}\rangle=0,\quad i=1,\dots,5,\qquad\mathbf{U}\in \mathcal{D}(\Omega,\otimes^2_0),
\end{equation} 
are the extension to ${^R\mathbf{g}}\in \mathcal{D}^{'}(\Omega,\otimes^0_2)$ of the concept of invariance of a tensor field under the $C^\infty$ (respectively $C^1$) transformations generated by $^R\mathbf{K}_i$, $i=1,2,3$ (respectively $i=4,5$).\footnote{Recall that a $C^r$ vector field generate in a natural way a $C^r$ flow.} Thus,
\begin{equation}
\mathfrak{Isom}(\Omega,{^R\mathbf{g}})=\{{^R\mathbf{K}}_i,\,i=1,\ldots,5\},
\end{equation}
where the only non zero Lie brackets are given by\footnote{The Lie bracket of $C^r$ vector fields, $r\geqslant 1$, is definable in the \textit{classical} sense, its local expression in coordinates given by  
\[
\overline{[\mathbf{V},\mathbf{W}]f}=(V^\alpha\partial_\alpha W^\beta-W^\alpha\partial_\alpha V^\beta)\partial_\beta f,
\]
for all $C^2$ functions f.}
\begin{equation}\label{noncommuting_algebra}
[{^R\mathbf{K}}_2,{^R\mathbf{K}}_4]={^R\mathbf{K}}_1,\quad [{^R\mathbf{K}}_3,{^R\mathbf{K}}_5]={^R\mathbf{K}}_1.
\end{equation}

\subsection{Conformal isometries of ${^R\mathbf{g}}$}  

Let us now look for the existence of conformal isometries of $^R\mathbf{g}$. Obviously, the homothety 
\begin{equation}\label{homothety}
u\rightarrow u,\quad v\rightarrow \xi^2 v,\quad x\rightarrow \xi x, \quad y\rightarrow \xi y,\qquad \xi=const.,
\end{equation}
is a conformal isometry of ${^R\mathbf{g}}$. This homothety is generated by the smooth vector field
\begin{equation}\label{conf_rosen_1}
{^R\mathbf{K}}_6= 2v\boldsymbol{\partial}_v + x\boldsymbol{\partial}_x + y\boldsymbol{\partial}_y.
\end{equation}
Indeed, if we compute $\mathcal{L}_{_{^R\mathbf{K}_6}}{^R\mathbf{g}}$ using (\ref{liederiv_a}) we find
\begin{equation}
\langle\mathcal{L}_{_{^R\mathbf{K}_6}}{^R\mathbf{g}},\mathbf{U}\rangle=\langle 2\,{^R\mathbf{g}},\mathbf{U}\rangle, \qquad\forall\, \mathbf{U}\in \mathcal{D}(\Omega,\otimes^2_0).
\end{equation} 
Note that for $\boldsymbol{\eta}$ given by (\ref{minkowski}) we have \cite{Zhang:2019gdm}
\begin{equation}
\mathcal{L}_{_{^R\mathbf{K}_6}}{\boldsymbol{\eta}}=2\,{\boldsymbol{\eta}}.
\end{equation}

Next, among the conformal Killing vector fields in the set $\mathfrak{Conf}(\mathbb{R}^4,\boldsymbol{\eta})$ with $\boldsymbol{\eta}$ given by (\ref{minkowski}), we have \cite{Zhang:2019gdm}
\begin{equation}
\mathcal{L}_{_{\mathbf{K}_K}}{\boldsymbol{\eta}}=2u\,{\boldsymbol{\eta}},\qquad 
\mathbf{K}_K= u^2\boldsymbol{\partial}_u+\frac{1}{2}\left(x^2+y^2\right)\boldsymbol{\partial}_v +ux\boldsymbol{\partial}_x +uy\boldsymbol{\partial}_y.
\end{equation}
The above suggests that we look for a conformal Killing vector field of ${^R\mathbf{g}}$ by solving the conformal Killing equation (\ref{CKV}) for the metric ${^R\mathbf{g}}$ with the conformal scalar $\psi=u$,  
\begin{equation}\label{conf_killing}
\mathcal{L}_{_{^R\mathbf{K}_7}}{^R\!\mathbf{g}}=2u\,{^R\!\mathbf{g}}.
\end{equation}
Admitting solutions of $C^1$ regularity across the surface $u=0$, we find 
\begin{equation}\label{conf_rosen_2}
^R\mathbf{K}_7= \begin{cases}
u^2\boldsymbol{\partial}_u+\frac{1}{2}\left(x^2+y^2\right)\boldsymbol{\partial}_v +ux\boldsymbol{\partial}_x +uy\boldsymbol{\partial}_y,\quad u<0,\\ \\
u^2\boldsymbol{\partial}_u+\frac{1}{2}\left(x^2+y^2\right)\boldsymbol{\partial}_v +({u}/({1+u}))x\boldsymbol{\partial}_x +({u}/({1-u}))y\boldsymbol{\partial}_y, \quad u\geqslant 0.
\end{cases}
\end{equation}
Note that the existence of this conformal Killing vector field of regularity $C^1$ is far from being trivial, it has a non-null component along $\boldsymbol{\partial}_u$ and the metric ${^R\mathbf{g}}$ is merely locally Lipchitz across $u=0$. Using (\ref{liederiv_b}) with $\boldsymbol{g}=\boldsymbol{\eta}$ we have
\begin{eqnarray}
\langle\mathcal{L}_{_{^R\mathbf{K}_7}}{^R\!\mathbf{g}},\mathbf{U}\rangle&=&
-\left\langle{^R\mathbf{g}},\,\mathcal{L}_{_{^R\mathbf{K}_7}} \mathbf{U} +
\frac{1}{2}(\boldsymbol{\eta}^{-1}|\mathcal{L}_{_{^R\mathbf{K}_7}}\boldsymbol{\eta}) \mathbf{U}\right\rangle\nonumber\\
&=&-\int_{-u_0}^0 du \int_{\mathbb{R}^3}dv\,dx\,dy \,\left[(u^2\partial_u +\frac{1}{2}(x^2+y^2)\partial_v +ux\partial_x +uy\partial_y +2u)\,(-U^{uv} - U^{vu}+ U^{xx} +U^{yy})\right] \nonumber \\
&&-\int_{0}^{u_0} du\int_{\mathbb{R}^3}dv\,dx\,dy\,\left[\frac{\!}{\!}\!\!-u^2\partial_u(U^{uv} + U^{vu}) +(1+u)^2u^2\partial_uU^{xx} +(1-u)^2u^2\partial_u U^{yy}\right.\nonumber\\ &&\left.\quad\,\,+\left(\frac{1}{2}(x^2+y^2)\partial_v +\frac{u}{1+u}(x\partial_x +1) + \frac{u}{1-u}(y\partial_y +1) +2u \right)(-U^{uv}-U^{vu} +(1+u)^2U^{xx}\right.\nonumber\\&& \left.\frac{\!}{\!}+(1-u)^2U^{yy})+ 2u(U^{uv} +U^{vu}) -2u(1+u)U^{xx} -2u(1-u)U^{yy}\right]\nonumber\\
&=&\int_{-u_0}^0 du \int_{\mathbb{R}^3}dv\,dx\,dy \,2u\,(-U^{uv} - U^{vu}+ U^{xx} +U^{yy}) \nonumber \\
&&+\int_{0}^{u_0} du\int_{\mathbb{R}^3}dv\,dx\,dy\,2u\,(-U^{uv} -U^{vu}+ (1+u)^2U^{xx} +(1-u)^2U^{yy}),\nonumber\\
&=& \langle 2u\,{^R\mathbf{g}},\mathbf{U}\rangle, \qquad \mathbf{U}\in \mathcal{D}(\Omega,\otimes^2_0)
\end{eqnarray}
where we have integrated by parts and used the fact that $\mathbf{U}$ is of compact support, 
from which (\ref{conf_killing}) follows in the sense of distributions.

As we have already mentioned, for a smooth planewave spacetime the maximum dimension of the algebra of all conformal Killing vector fields is seven. For $\mathcal{L}_{_{\mathbf{K}}}{^R\!\mathbf{g}}\in \mathcal{D}^{'1}(\Omega,\otimes^0_2)$, we find that $\mathfrak{Conf}({\Omega},{^R\mathbf{g}})$ is seven-dimensional: the isometry algebra $\mathfrak{Isom}(\Omega,{^R\mathbf{g}})$ is five-dimensional, there is a homothetic Killing vector field ${^R\mathbf{K}}_6$ and one conformal Killing vector field ${^R\mathbf{K}}_7$. Hence, for ${^R\mathbf{g}}\in \mathcal{D}^{'}(\Omega,\otimes^0_2)$ we have
\begin{equation}
\mathfrak{Conf}({\Omega},{^R\mathbf{g}})=\{{^R\mathbf{K}}_i,\,i=1,\ldots,7\},
\end{equation}
where
\begin{equation}
[{^R\mathbf{K}}_2,{^R\mathbf{K}}_4]={^R\mathbf{K}}_1,\qquad [{^R\mathbf{K}}_3,{^R\mathbf{K}}_5]={^R\mathbf{K}}_1,
\end{equation} 
\begin{equation}
[{^R\mathbf{K}}_1,{^R\mathbf{K}}_6]=2\,{^R\mathbf{K}}_1,\quad [{^R\mathbf{K}}_i, {^R\mathbf{K}}_6]={^R\mathbf{K}}_i,\quad i=2, \dots, 5
\end{equation}
and
\begin{equation}
[{^R\mathbf{K}}_2, {^R\mathbf{K}}_7]={^R\mathbf{K}}_4,,\quad [{^R\mathbf{K}}_3, {^R\mathbf{K}}_7]={^R\mathbf{K}}_5,
\end{equation}
all the other Lie brackets are zero (see footnote 7).

\section{Curvature collineations of the impulsive plane wave}\label{CC_section}

For a spacetime $(\mathcal{M},\mathbf{g})$, where $\mathbf{g}$ is
a $C^{\infty}$ metric tensor and $\mathbf{Riem}$ its curvature tensor field, a curvature collineation $\mathbf{K}$ is defined as the solution to \cite{Katzin69}
\begin{equation}\label{CC}
\mathcal{L}_{_{\mathbf{K}}}\mathbf{Riem}=0.
\end{equation}
The associated vector fields $\mathbf{K}$ may constitute an infinite-dimensional vector space and the set of all smooth curvature collineations on $\mathcal{M}$ is a Lie algebra with $\mathfrak{Isom}(\mathcal{M},\mathbf{g})$ a subset of this one. The importance and close connection of curvature collineations with conservation laws has been stated some time ago \cite{Katzin69,Katzin72}. For smooth pp-wave metrics in the Brinkmann form (\ref{pp_wave}), the general form of the vector field $\mathbf{K}$ that satisfies (\ref{CC}) was found in Ref.\cite{Aichelburg70}.  

Now, we turn our attention to the curvature $\mathbf{Riem}$ of the metric ${^R\mathbf{g}}$ (\ref{pprosen_2}). As we have already mentioned, although (locally Lipschitz) continuous, ${^R\mathbf{g}}$ is not a $C^\infty$ metric and therefore lies beyond the reach of classical Lorentzian geometry. We shall pursued a distributional approach to compute its curvature. The most general class of metrics for which the Riemann curvature tensor and its contractions can be interpreted rigorously as distributions has been identified by Geroch and Traschen in Ref.\cite{geroch} (for a geometric coordinate free description see Ref.\cite{LeFloch:2007tv}). Metrics for thin shells \cite{israel,barrabes} are included into this class, although metrics with distributional coefficients, as the one given by (\ref{pp_brinkmann}), are not. 

It can be readily verified that ${^R\mathbf{g}}$ (\ref{pprosen_2}) on $\Omega$ lies within the Geroch-Traschen class of metrics, which is ${^R\mathbf{g}}$ locally in $H^1\cap L^\infty$ \cite{geroch, LeFloch:2007tv}. For the sake of brevity we shall skip the proof and simply sketch the calculation of its curvature $\mathbf{Riem}$. Let $\mathbf{U}$ be a test $(3,1)$-tensor with compact support in $\Omega$. Let $\boldsymbol{\eta}$ be the four-dimensional Minkowski
metric tensor in light-cone coordinates (\ref{minkowski}) and let ${^{\boldsymbol{\eta}}\nabla}$ be the covariant derivative in $\boldsymbol{\eta}$. The Riemann curvature is the distribution-valued tensor $\mathbf{Riem}$ of order $\leqslant 1$ given by
\begin{equation}\label{riemann_1}
\langle \mathbf{Riem},\mathbf{U}\rangle\equiv \int_{\varphi(\Omega)} \left(-2C^{\delta}_{\,\,\gamma [ \alpha}{^{\boldsymbol{\eta}}\nabla}_{\beta ]} U_{\,\,\,\,\,\,\,\,\,\,\delta}^{\alpha\beta\gamma} +2C^\delta_{\,\,\epsilon[\alpha}C^\epsilon_{\,\,\beta]\gamma}U_{\,\,\,\,\,\,\,\,\,\,\delta}^{\alpha\beta\gamma}\right)\boldsymbol{\omega}_{\boldsymbol{\eta}}, 
\end{equation}
where the Christoffel tensor $\mathbf{C}$ is the locally square-integrable tensor field given by 
\begin{equation}
\mathbf{C}=\begin{cases}
0,\qquad u<0;\\
\\
{(1+u)^{-1}}\left(\boldsymbol{\partial}_x\otimes\mathbf{d}x\otimes\mathbf{d}u +\boldsymbol{\partial}_x\otimes\mathbf{d}u\otimes\mathbf{d}x\right)-{(1-u)^{-1}}\left(\boldsymbol{\partial}_y\otimes\mathbf{d}y\otimes\mathbf{d}u +\boldsymbol{\partial}_y\otimes\mathbf{d}u\otimes\mathbf{d}y\right)\\+(1+u)\boldsymbol{\partial}_v\otimes\mathbf{d}x\otimes\mathbf{d}x 
-(1-u)\boldsymbol{\partial}_v\otimes\mathbf{d}y\otimes\mathbf{d}y,\qquad u>0.
\end{cases} 
\end{equation}
After integration by parts and using the fact that $\mathbf{U}$ is of compact support, (\ref{riemann_1}) gives
\[
\langle \mathbf{Riem},\mathbf{U}\rangle= \int_{\mathbb{R}^3}dv\,dx\,dy\, \left.\left(U_{\,\,\,\,\,\,\,\,\,\,x}^{xuu}-U_{\,\,\,\,\,\,\,\,\,\,x}^{uxu} + U_{\,\,\,\,\,\,\,\,\,\,y}^{uyu}-U_{\,\,\,\,\,\,\,\,\,\,y}^{yuu}+ U_{\,\,\,\,\,\,\,\,\,\,v}^{xux} -U_{\,\,\,\,\,\,\,\,\,\,v}^{uxx} + U_{\,\,\,\,\,\,\,\,\,\,v}^{uyy} -U_{\,\,\,\,\,\,\,\,\,\,v}^{yuy}\right)\right|_{u=0},
\]     
from which we read off the distribution-valued curvature tensor $\mathbf{Riem}$ of ${^R\mathbf{g}}$,
\begin{equation}\label{riemann_curv}
\mathbf{Riem}=\delta(u)\boldsymbol{\mathcal{T}},
\end{equation}
where $\boldsymbol{\mathcal{T}}$ is the smooth $(1,3)$-tensor field given by
\begin{eqnarray}\label{Riem_structure}
\boldsymbol{\mathcal{T}}=&&\left(\mathbf{d}x\otimes\mathbf{d}u\otimes\mathbf{d}u\otimes \boldsymbol{\partial}_x - \mathbf{d}u\otimes\mathbf{d}x\otimes\mathbf{d}u\otimes \boldsymbol{\partial}_x\, + \mathbf{d}u\otimes\mathbf{d}y\otimes\mathbf{d}u \otimes\boldsymbol{\partial}_y- \mathbf{d}y\otimes\mathbf{d}u\otimes\mathbf{d}u\otimes\boldsymbol{\partial}_y\right.\nonumber\\&& \left.+ \mathbf{d}x\otimes\mathbf{d}u\otimes\mathbf{d}x\otimes\boldsymbol{\partial}_v - \mathbf{d}u\otimes\mathbf{d}x\otimes\mathbf{d}x \otimes\boldsymbol{\partial}_v+ \mathbf{d}u\otimes\mathbf{d}y\otimes\mathbf{d}y \otimes\boldsymbol{\partial}_v- \mathbf{d}y\otimes\mathbf{d}u\otimes\mathbf{d}y\otimes\boldsymbol{\partial}_v\right).
\end{eqnarray}
It follows that $\mathbf{Riem}$ is supported on the 3-dimensional infinite plane $\{(u,v,x,y)\in \Omega \,|\, u=0\}$ and the spacetime is flat away from this hypersurface.

Now, for $\mathbf{Riem}$ given by (\ref{riemann_curv}), we can make use of (\ref{liederiv_b}) to evaluate $\mathcal{L}_{_{\mathbf{K}}}\mathbf{Riem}$ along the vector fields in $\mathfrak{Isom}(\Omega,{^R\mathbf{g}})$. As before, let $\boldsymbol{\eta}$ be the four-dimensional Minkowski
metric tensor (\ref{minkowski}). Let $\mathbf{U}$ a test $(3,1)$-tensor with compact support in $\Omega$. For the Killing vector $^R\mathbf{K}_1=\boldsymbol{\partial}_v$ we have 
\begin{eqnarray}
\langle\mathcal{L}_{_{^R\mathbf{K}_1}}\mathbf{Riem},\mathbf{U}\rangle\equiv&&
-\langle \mathbf{Riem}, \mathcal{L}_{_{^R\mathbf{K}_1}}\mathbf{U} + \frac{1}{2}(\boldsymbol{\eta}^{-1}|\mathcal{L}_{_{^R\mathbf{K}_1}}\boldsymbol{\eta})\mathbf{U}\rangle\nonumber\\
=&&-\int_{\mathbb{R}^3}dv\,dx\,dy\,\,\partial_v\left.\left(\mathcal{T}_{\alpha\beta\gamma}^{\,\,\,\,\,\,\,\,\,\,\delta}U^{\alpha\beta\gamma}_{\,\,\,\,\,\,\,\,\,\,\,\delta}\right)\right|_{u=0},\nonumber
\end{eqnarray}
where $\mathcal{T}_{\alpha\beta\gamma}^{\,\,\,\,\,\,\,\,\,\,\delta}$ are the components of the smooth $(1,3)$-tensor field $\boldsymbol{\mathcal{T}}$ given by (\ref{Riem_structure}).
Using the fact that $\mathbf{U}$ is of compact support, we find
\begin{equation}
\langle\mathcal{L}_{_{^R\mathbf{K}_1}}\mathbf{Riem},\mathbf{U}\rangle=0,\qquad\forall\, \mathbf{U}\in \mathcal{D}(\Omega,\otimes^3_1).
\end{equation}
Analogous calculations show that $\mathcal{L}_{_{^R\mathbf{K}_2}}\mathbf{Riem}$ and $\mathcal{L}_{_{^R\mathbf{K}_3}}\mathbf{Riem}$ are zero distributions on $\Omega$.

Next, consider the $C^1$ Killing vector field ${^R\mathbf{K}}_4$. Let $\boldsymbol{g}$ be any smooth metric in $\Omega$. Note that $\mathcal{L}_{_{^R\mathbf{K}_4}}\mathbf{U}$ and $(\boldsymbol{g}^{-1}|\mathcal{L}_{_{^R\mathbf{K}_4}}\boldsymbol{g})\mathbf{U}$ are $C^0$ $(3,1)$-tensor fields with compact support on $\Omega$ if $\mathbf{U}\in \mathcal{D}^{1}(\Omega,\otimes^3_1)$. Since $\mathbf{Riem}$, as given by (\ref{riemann_curv}), is a distribution-valued $(1,3)$-tensor field of order 0, it follows that the quantity $\langle \mathbf{Riem}, \mathcal{L}_{_{^R\mathbf{K}_4}}\mathbf{U} + (1/2)(\boldsymbol{g}^{-1}|\mathcal{L}_{_{^R\mathbf{K}_4}}\boldsymbol{g})\mathbf{U}\rangle$ is defined. Hence, a well defined distributional meaning can be given to $\mathcal{L}_{_{^R\mathbf{K}_4}} \mathbf{Riem}$ {\it via} (\ref{liederiv_b}). Let $\boldsymbol{\eta}$ be the four-dimensional Minkowski metric tensor (\ref{minkowski}). For the Killing vector $^R\mathbf{K}_4$ and a test tensor $\mathbf{U}\in \mathcal{D}(\Omega,\otimes^3_1)$ we have
\begin{eqnarray}
\langle \mathbf{Riem}, \mathcal{L}_{_{^R\mathbf{K}_4}}\mathbf{U} + \frac{1}{2}(\boldsymbol{\eta}^{-1}|\mathcal{L}_{_{^R\mathbf{K}_4}}\boldsymbol{\eta})\mathbf{U}\rangle
&=&\int_{\mathbb{R}^3}dv\,dx\,dy\,\left[x\partial_v\left(\mathcal{T}_{\alpha\beta\gamma}^{\,\,\,\,\,\,\,\,\,\,\delta}U^{\alpha\beta\gamma}_{\,\,\,\,\,\,\,\,\,\,\,\delta}\right)\right.\nonumber\\&&-\left.\left.\left(\mathcal{T}_{x\beta\gamma}^{\,\,\,\,\,\,\,\,\,\,\delta}U^{u\beta\gamma}_{\,\,\,\,\,\,\,\,\,\,\,\delta}+\mathcal{T}_{\alpha x\gamma}^{\,\,\,\,\,\,\,\,\,\,\delta}U^{\alpha u\gamma}_{\,\,\,\,\,\,\,\,\,\,\,\delta}+\mathcal{T}_{\alpha\beta x}^{\,\,\,\,\,\,\,\,\,\,\delta}U^{\alpha\beta u}_{\,\,\,\,\,\,\,\,\,\,\,\delta}-\mathcal{T}_{\alpha\beta\gamma}^{\,\,\,\,\,\,\,\,\,\,x}U^{\alpha\beta\gamma}_{\,\,\,\,\,\,\,\,\,\,\,v}\right)\right]\right|_{u=0}.\nonumber
\end{eqnarray}
Now, from (\ref{Riem_structure}) we find
\[
\mathcal{T}_{x\beta\gamma}^{\,\,\,\,\,\,\,\,\,\,\delta}U^{u\beta\gamma}_{\,\,\,\,\,\,\,\,\,\,\,\delta}+\mathcal{T}_{\alpha x\gamma}^{\,\,\,\,\,\,\,\,\,\,\delta}U^{\alpha u\gamma}_{\,\,\,\,\,\,\,\,\,\,\,\delta}+\mathcal{T}_{\alpha\beta x}^{\,\,\,\,\,\,\,\,\,\,\delta}U^{\alpha\beta u}_{\,\,\,\,\,\,\,\,\,\,\,\delta}-\mathcal{T}_{\alpha\beta\gamma}^{\,\,\,\,\,\,\,\,\,\,x}U^{\alpha\beta\gamma}_{\,\,\,\,\,\,\,\,\,\,\,v}=0.
\]
Thus, 
\begin{equation}
\langle \mathbf{Riem}, \mathcal{L}_{_{^R\mathbf{K}_4}}\mathbf{U} + \frac{1}{2}(\boldsymbol{\eta}^{-1}|\mathcal{L}_{_{^R\mathbf{K}_4}}\boldsymbol{\eta})\mathbf{U}\rangle=-\int_{\mathbb{R}^3}dv\,dx\,dy\,\,\left.\partial_v\left(x\,\mathcal{T}_{\alpha\beta\gamma}^{\,\,\,\,\,\,\,\,\,\,\delta}U^{\alpha\beta\gamma}_{\,\,\,\,\,\,\,\,\,\,\,\delta}\right)\right|_{u=0}=0,
\end{equation}
where we have use the fact that $\mathbf{U}$ is of compact support. Analogous calculations show that $
\langle\mathcal{L}_{_{^R\mathbf{K}_5}}\mathbf{Riem},\mathbf{U}\rangle=0$, $\mathbf{U}\in \mathcal{D}(\Omega,\otimes^3_1)$. Hence, $\mathcal{L}_{_{^R\mathbf{K}_4}}\mathbf{Riem}=0$ and $\mathcal{L}_{_{^R\mathbf{K}_5}}\mathbf{Riem}=0$ in the sense of distributions. We see that, as in smooth spacetimes, all the Killing vector fields are special curvature collineations. However, it should be stressed that for the metric $^R\mathbf{g}$ (\ref{pprosen_2}) and its curvature $\mathbf{Riem}$ (\ref{riemann_curv}) this fact obviously cannot be proved outside the distributional setting. 

Next, since the spacetime $(\Omega,{^R\mathbf{g}})$ is Ricci-flat, $\mathbf{Ric}=0$, the curvature $\mathbf{Riem}$ is equal to the conformally invariant curvature $\mathbf{Weyl}$. It follows that every conformal isometry is a curvature collineation. In fact, for $\mathbf{K}_6$ (\ref{conf_rosen_1}) we have
           \begin{eqnarray*}
           \langle \mathcal{L}_{^R\mathbf{K}_6}\mathbf{Riem},\mathbf{U}\rangle=&&-\langle \mathbf{Riem}, \mathcal{L}_{^R\mathbf{K}_6}\mathbf{U} +\frac{1}{2}(\boldsymbol{\eta}^{-1}|\mathcal{L}_{_{^R\mathbf{K}_6}}\boldsymbol{\eta})\mathbf{U}\rangle\\=&& -\int_{\mathbb{R}^3}dv\,dx\,dy\,\left[\left(2v\partial_v + x\partial_x + y\partial_y\right)\left(\mathcal{T}_{\alpha\beta\gamma}^{\,\,\,\,\,\,\,\,\,\,\delta}U^{\alpha\beta\gamma}_{\,\,\,\,\,\,\,\,\,\,\delta}\right)\right.\\ &&- \left.\left(\mathcal{T}_{x\beta\gamma}^{\,\,\,\,\,\,\,\,\,\,\delta}U^{x\beta\gamma}_{\,\,\,\,\,\,\,\,\,\,\delta}+\mathcal{T}_{y\beta\gamma}^{\,\,\,\,\,\,\,\,\,\,\delta}U^{y\beta\gamma}_{\,\,\,\,\,\,\,\,\,\,\delta}+\mathcal{T}_{\alpha x\gamma}^{\,\,\,\,\,\,\,\,\,\,\delta}U^{\alpha x\gamma}_{\,\,\,\,\,\,\,\,\,\,\delta}+\mathcal{T}_{\alpha y\gamma}^{\,\,\,\,\,\,\,\,\,\,\delta}U^{\alpha y\gamma}_{\,\,\,\,\,\,\,\,\,\,\delta}+\mathcal{T}_{\alpha\beta x}^{\,\,\,\,\,\,\,\,\,\,\delta}U^{\alpha\beta x}_{\,\,\,\,\,\,\,\,\,\,\delta}+\mathcal{T}_{\alpha\beta y}^{\,\,\,\,\,\,\,\,\,\,\delta}U^{\alpha\beta y}_{\,\,\,\,\,\,\,\,\,\,\delta}\right.\right.\nonumber\\&&\left.\left.\left.-\mathcal{T}_{\alpha\beta\gamma}^{\,\,\,\,\,\,\,\,\,\,x}U^{\alpha\beta\gamma}_{\,\,\,\,\,\,\,\,\,\,x}-\mathcal{T}_{\alpha\beta\gamma}^{\,\,\,\,\,\,\,\,\,\,y}U^{\alpha\beta\gamma}_{\,\,\,\,\,\,\,\,\,\,y}-2\mathcal{T}_{\alpha\beta\gamma}^{\,\,\,\,\,\,\,\,\,\,v}U^{\alpha\beta\gamma}_{\,\,\,\,\,\,\,\,\,\,v} \right) + 4\mathcal{T}_{\alpha\beta\gamma}^{\,\,\,\,\,\,\,\,\,\,\delta}U^{\alpha\beta\gamma}_{\,\,\,\,\,\,\,\,\,\,\delta}\right]\right|_{u=0}\nonumber\\=&& -\int_{\mathbb{R}^3}dv\,dx\,dy\,\left.\left[\left(2v\partial_v + x\partial_x + y\partial_y\right)\left(\mathcal{T}_{\alpha\beta\gamma}^{\,\,\,\,\,\,\,\,\,\,\delta}U^{\alpha\beta\gamma}_{\,\,\,\,\,\,\,\,\,\,\delta}\right) + 4\mathcal{T}_{\alpha\beta\gamma}^{\,\,\,\,\,\,\,\,\,\,\delta}U^{\alpha\beta\gamma}_{\,\,\,\,\,\,\,\,\,\,\delta}\right]\right|_{u=0}=0,\quad\forall\, \mathbf{U}\in \mathcal{D}(\Omega,\otimes^3_1),
 \end{eqnarray*} 
 where we have made use of the explicit form of the components $\mathcal{T}_{\alpha\beta\gamma}^{\,\,\,\,\,\,\,\,\,\,\delta}$, integrated by parts and used the fact that $\mathbf{U}$ is of compact support. On the other hand, for $\mathbf{K}_7$ (\ref{conf_rosen_2}) we have 
  \begin{eqnarray*}
           \langle \mathcal{L}_{^R\mathbf{K}_7}\mathbf{Riem},\mathbf{U}\rangle=&&-\langle \mathbf{Riem}, \mathcal{L}_{^R\mathbf{K}_7}\mathbf{U} +\frac{1}{2}(\boldsymbol{\eta}^{-1}|\mathcal{L}_{_{^R\mathbf{K}_7}}\boldsymbol{\eta})\mathbf{U}\rangle\\=&& -\int_{\mathbb{R}^3}dv\,dx\,dy\,\left[\frac{1}{2}(x^2+y^2)\partial_v\left(\mathcal{T}_{\alpha\beta\gamma}^{\,\,\,\,\,\,\,\,\,\,\delta}U^{\alpha\beta\gamma}_{\,\,\,\,\,\,\,\,\,\,\delta}\right)\right.\\ &&- x\left(\mathcal{T}_{x\beta\gamma}^{\,\,\,\,\,\,\,\,\,\,\delta}U^{u\beta\gamma}_{\,\,\,\,\,\,\,\,\,\,\delta}+\mathcal{T}_{\alpha x\gamma}^{\,\,\,\,\,\,\,\,\,\,\delta}U^{\alpha u\gamma}_{\,\,\,\,\,\,\,\,\,\,\delta}+\mathcal{T}_{\alpha\beta x}^{\,\,\,\,\,\,\,\,\,\,\delta}U^{\alpha\beta u}_{\,\,\,\,\,\,\,\,\,\,\delta}-\mathcal{T}_{\alpha\beta\gamma}^{\,\,\,\,\,\,\,\,\,\,x}U^{\alpha\beta\gamma}_{\,\,\,\,\,\,\,\,\,\,v}\right)\nonumber\\&&- \left.\left.y\left(\mathcal{T}_{y\beta\gamma}^{\,\,\,\,\,\,\,\,\,\,\delta}U^{u\beta\gamma}_{\,\,\,\,\,\,\,\,\,\,\delta}+\mathcal{T}_{\alpha y\gamma}^{\,\,\,\,\,\,\,\,\,\,\delta}U^{\alpha u\gamma}_{\,\,\,\,\,\,\,\,\,\,\delta}+\mathcal{T}_{\alpha\beta y}^{\,\,\,\,\,\,\,\,\,\,\delta}U^{\alpha\beta u}_{\,\,\,\,\,\,\,\,\,\,\delta}-\mathcal{T}_{\alpha\beta\gamma}^{\,\,\,\,\,\,\,\,\,\,y}U^{\alpha\beta\gamma}_{\,\,\,\,\,\,\,\,\,\,v}\right) \right]\right|_{u=0}\nonumber\\=&& -\int_{\mathbb{R}^3}dv\,dx\,dy\,\left.\partial_v\left[\frac{1}{2}(x^2+y^2) \left(\mathcal{T}_{\alpha\beta\gamma}^{\,\,\,\,\,\,\,\,\,\,\delta}U^{\alpha\beta\gamma}_{\,\,\,\,\,\,\,\,\,\,\delta}\right)\right]\right|_{u=0}=0,\quad\forall\, \mathbf{U}\in \mathcal{D}(\Omega,\otimes^3_1),
 \end{eqnarray*}
where we have made use of the explicit form of the components $\mathcal{T}_{\alpha\beta\gamma}^{\,\,\,\,\,\,\,\,\,\,\delta}$ and the fact that $\mathbf{U}$ is of compact support. Thus, $\mathcal{L}_{_{^R\mathbf{K}_6}} \mathbf{Riem}=0$ and $\mathcal{L}_{_{^R\mathbf{K}_7}} \mathbf{Riem}=0$ in the sense of distributions. 

We have so far showed that the vector fields in the conformal algebra $\mathfrak{Conf}({\Omega},{^R\mathbf{g}})$ are curvature collineations of $\mathbf{Riem}$ (\ref{riemann_curv}).
Now we ask for the general form of a curvature collineation of $\mathbf{Riem}$ (\ref{riemann_curv}), i.e. for the general form of a (smooth or $C^{k}$, $k \geqslant 1$) vector field $\mathbf{K}$ such  that (\ref{CC}) holds for $\mathbf{Riem}$ (\ref{riemann_curv}) in the sense of distributions,
\begin{equation}\label{dist_CC_1}
\langle\mathcal{L}_{_{\mathbf{K}}}\mathbf{Riem},\mathbf{U}\rangle=0,\qquad \forall \mathbf{U}\in \mathcal{D}(\Omega,\otimes^3_1).
\end{equation}
From the definition of the distributional Lie derivative (\ref{liederiv_b}), (\ref{dist_CC_1}) implies 
\begin{equation}\label{dist_CC_2}
\langle\mathbf{Riem},\mathcal{L}_{_{^R\mathbf{K}}}\mathbf{U} + \frac{1}{2}(\boldsymbol{g}^{-1}|\mathcal{L}_{_{^R\mathbf{K}}}\boldsymbol{g})\mathbf{U}\rangle=0,\qquad \forall \mathbf{U}\in \mathcal{D}(\Omega,\otimes^3_1).
\end{equation}
The results of Ref.\cite{Aichelburg70}\footnote{For smooth pp wave spacetimes, it can be shown that the curvature tensor field in Brinkmann coordinates \cite{Aichelburg70} exhibits the same pattern of non-zero components that occurs in Balwin-Jeffery-Rosen coordinates.} and the explicit form of the elements of $\mathfrak{Conf}({\Omega},{^R\mathbf{g}})$ suggest the ansatz 
\begin{eqnarray}\label{CC_2_K_ansatz}
{^R\mathbf{K}}&=&\Phi(u)\boldsymbol{\partial}_u + \left[ \frac{1}{2}\Lambda(u)\left(x^2 + y^2\right) +\Psi_x(u)x+ \Psi_y(u)y +\Upsilon(u)v +\omega(u)\right]\boldsymbol{\partial}_v\nonumber\\&& +  \left[\Xi_x(u)x +\chi_x(u)\right]\boldsymbol{\partial}_x + \left[\Xi_y(u)y +\chi_y(u)\right]\boldsymbol{\partial}_y, 
\end{eqnarray}
where $\Phi(u)$, $\Lambda(u)$, $\Psi_x(u)$, $\Psi_y(u)$, $\Upsilon(u)$, $\omega(u)$, $\Xi_x(u)$, $\Xi_y(u)$, $\chi_x(u)$ and $\chi_y(u)$ are smooth or $C^{k}$ functions of $u$ with $k \geqslant 1$. 
Hence, for $\mathbf{Riem}$ given by (\ref{riemann_curv}) and $\mathbf{K}$ given by (\ref{CC_2_K_ansatz}), we will demand that (\ref{dist_CC_2}) be satisfied.
The detailed calculation, although straightforward, is too lengthy to be exhibited. We find that (\ref{dist_CC_2}) is satisfied whenever 
\begin{equation}\label{conditions}
\Phi(0)=0=\Phi'(0),\quad
\Lambda(0)=\Xi'_x(0)=\Xi'_y(0),\quad\Psi_x(0)=\chi'_x(0),\quad \Psi_y(0)=\chi'_y(0),\quad\Upsilon(0)=2\,\Xi_x(0)=2\,\Xi_y(0),
\end{equation}
where $'$ denotes the derivative with respect to $u$. Thus, (\ref{conditions})   
are additional conditions which must be imposed at $u=0$ on these otherwise arbitrary $C^{k}$ functions, $k \geqslant 1$.  

 As expected, for particular values of these functions we recover the isometry algebra 
 $\mathfrak{Isom}({\Omega},{^R\mathbf{g}})$ as well also the conformal isometry algebra 
 $\mathfrak{Conf}({\Omega},{^R\mathbf{g}})$. Of course, the set of all smooth curvature 
 collineations of the form (\ref{CC_2_K_ansatz}) is also a Lie algebra. Finally, note that 
 BMS-like supertranslations, which appear as isometries of the pullback of the metric tensor 
 to the hypersurface $u=0$ where $\mathbf{Riem}$ is supported, are not curvature collineations.

\section{Conformal isometries of the impulsive plane wave in Brinkmann form}\label{brinkmann_form}

Geometric symmetries of pp-wave spacetimes are usually discussed in Brinkmann 
coordinates (see 
Ref.\cite{tupper} and references therein). Using the Brinkmann form (\ref{pp_wave}) of the 
metric, it has been shown that all impulsive pp-waves admit at least a 3-parameter group of 
local $C^\infty$ isometries \cite{aichelburg95}, while for certain classes the existence of   
additional non-smooth Killing vectors was put forward in Ref.\cite{aichelburg96}. In particular, for the impulsive plane wave $^B\mathbf{g}$ (\ref{pp_brinkmann}), it is found that 
\begin{equation}\label{Killing_eq_brinkmann}
\mathcal{L}_{_{^B\mathbf{K}_i}}{^B\mathbf{g}}=0,\quad i=1,\ldots,5;
\end{equation}
where \cite{aichelburg96}
\[
^B\mathbf{K}_1 = \boldsymbol{\partial} _V, \quad
^B\mathbf{K}_2 = \boldsymbol{\partial} _X +\Theta_{U}^+{^B\mathbf{K}}_4, \quad
^B\mathbf{K}_3 = \boldsymbol{\partial} _Y - \Theta_{U}^+{^B\mathbf{K}}_5,
\]
\begin{equation}\label{killings_brinkmann}
^B\mathbf{K}_4 =  X\boldsymbol{\partial}_V + U\boldsymbol{\partial}_X, \quad
^B\mathbf{K}_5 =   Y\boldsymbol{\partial}_V + U\boldsymbol{\partial}_Y,
\end{equation}
which satisfy 
 \begin{equation}
[{^B\mathbf{K}}_2,{^B\mathbf{K}}_4]={^B\mathbf{K}}_1,\qquad [{^B\mathbf{K}}_3,{^B\mathbf{K}}_5]={^B\mathbf{K}}_1,
\end{equation}
all the other Lie brackets being zero. The 3-dimensional abelian subalgebra spanned by the 
smooth Killing vector fields $\{^B\mathbf{K}_1,^B\mathbf{K}_4,^B\mathbf{K}_5\}$ is common 
to all impulsive pp-waves \cite{aichelburg95}. The two additional isometries, associated to the 
regularly discontinuous (across the surface $U=0$) vector fields $^B\mathbf{K}_2$ and 
$^B\mathbf{K}_3$, occur for the specific form of the profile (\ref{plus_polarization}) and were obtained as solutions of (\ref{Killing_eq_brinkmann}) \textit{via} a regularization scheme \cite{aichelburg96}. Remarkably, as was already mentioned, the Killing vector fields 
(\ref{ksuavespp3}) were obtained in Ref.\cite{aichelburg96} as the formal pullback of 
the set $\{{^B\mathbf{K}_i},\, i=1,\ldots, 5\}$ under the discontinuous coordinate change 
(\ref{trans}). 

Before proceeding to determine the conformal Killing vector fields of ${^B\mathbf{g}}$ (\ref{pp_brinkmann}), we briefly comment on its Killing vector fields (\ref{killings_brinkmann}) within our approach. Indeed, Eq.(\ref{liederiv_a}) can be used to compute the Lie derivatives of $^B\mathbf{g}$ (\ref{pp_brinkmann}) with respect to those Killing vector fields which are smooth. Let $\mathbf{U}$ be a test $(2,0)$-tensor with compact support on an open subset of $\mathbb{R}^4$. Choose $\boldsymbol{g}=\boldsymbol{\eta}$, with $\boldsymbol{\eta}$ the four-dimensional Minkowski
metric tensor in light-cone coordinates $(U,V,X,Y)$ 
\begin{equation}
\boldsymbol{\eta}= -\mathbf{d}U\otimes \mathbf{d}V - \mathbf{d}V\otimes
\mathbf{d}U + \mathbf{d}X\otimes \mathbf{d}X + \mathbf{d}Y\otimes
\mathbf{d}Y.\label{minkowski_B}
\end{equation}
As expected, we find 
\begin{equation}\label{k4_brinkmann}
\langle\mathcal{L}_{_{^B\mathbf{K}_i}}{^B\mathbf{g}},\mathbf{U}\rangle=0, \quad i=1,4,5, \qquad\forall\, \mathbf{U}\in \mathcal{D}(\mathbb{R}^4,\otimes^2_0).
\end{equation}
On the other hand, being the vector fields $^B\mathbf{K}_2$ and $^B\mathbf{K}_3$ of a regularity below $C^1$, the Lie derivatives of the distribution-valued metric $^B\mathbf{g}$ with respect to $^B\mathbf{K}_2$ or $^B\mathbf{K}_3$ can not be definable via (\ref{liederiv_b}).  

We now consider the conformal isometries of ${^B\mathbf{g}}$. For a pp-wave metric $\mathbf{g}$ as given by (\ref{pp_wave}), with $A(U,X,Y)$ given by (\ref{plane_wave},\ref{plus_polarization}), there exists a homothety of the spacetime 
\[
U\rightarrow U,\quad V\rightarrow \xi^2 V,\quad X\rightarrow \xi X, \quad Y\rightarrow \xi Y,\qquad \xi=const.,
\]
By simple inspection, an analogous homothety for the impulsive pp-wave space-time $(\mathbb{R}^4, {^B\mathbf{g}})$ also exists. In fact, using (\ref{liederiv_a}), it is straightforward to show that the smooth vector field $^B\mathbf{K}_6$, as given by 
\begin{equation}\label{conf_brinkmann_1}
^B\mathbf{K}_6=2V\boldsymbol{\partial}_V+X\boldsymbol{\partial}_X + Y\boldsymbol{\partial}_Y, 
\end{equation}
is a homothetic Killing vector of the distributional metric $^B\mathbf{g}$. Noting that $^B\mathbf{g}$ (\ref{pp_brinkmann}) can be written as
\begin{equation}\label{pp_b_2}
^B\mathbf{g}=\delta_{(U)}\!\left(X^2-Y^2\right)\mathbf{d}U\otimes\mathbf{d}U + \boldsymbol{\eta},
\end{equation}
where $\boldsymbol{\eta}$ is given by (\ref{minkowski_B}), and that 
\begin{equation}
\mathcal{L}_{_{^B\mathbf{K}_6}}{\boldsymbol{\eta}}=2\boldsymbol{\eta} \quad \Longrightarrow\quad
\langle \mathcal{L}_{_{^B\mathbf{K}_6}}{\boldsymbol{\eta}},\mathbf{U}\rangle=2 \langle \boldsymbol{\eta}, \mathbf{U}\rangle,\qquad \forall\, \mathbf{U}\in \mathcal{D}(\mathbb{R}^4,\otimes^2_0),
\end{equation}
we see that we need to calculate 
\[
\langle\mathcal{L}_{_{^B\mathbf{K}_6}}\delta_{(U)}\!\left(X^2-Y^2\right)\mathbf{d}U\otimes\mathbf{d}U,\mathbf{U}\rangle=-\left\langle \delta_{(U)}\!\left(X^2-Y^2\right)\mathbf{d}U\otimes\mathbf{d}U, \mathcal{L}_{_{^B\mathbf{K}_6}} \mathbf{U} +
\frac{1}{2}(\boldsymbol{g}^{-1}|\mathcal{L}_{_{{^B\mathbf{K}}_6}}\boldsymbol{g})\mathbf{U}\right\rangle.
\]
For $\boldsymbol{g}=\boldsymbol{\eta}$, we have $\frac{1}{2}(\boldsymbol{\eta}^{-1}|\mathcal{L}_{_{{^B\mathbf{K}}_6}}\boldsymbol{\eta})= ({^{\boldsymbol{\eta}}\nabla}\cdot{^B\mathbf{K}}_6)=4$ and we find
\begin{eqnarray*}
\left\langle \delta_{(U)}\!\left(X^2-Y^2\right)\mathbf{d}U\otimes\mathbf{d}U, \mathcal{L}_{_{^B\mathbf{K}_6}} \mathbf{U} + \frac{1}{2}(\boldsymbol{\eta}^{-1}|\mathcal{L}_{_{{^B\mathbf{K}}_6}}\boldsymbol{\eta})\mathbf{U}\right\rangle
&=&\int_{U=0}dV dX dY\,\left(X^2-Y^2\right)\!\left(2V\partial_V +X\partial_X +Y \partial_Y+4\right){U}^{UU}\\
&=&-\int_{U=0}dV dX dY \,2\left(X^2-Y^2\right){U}^{UU},  
\end{eqnarray*}
where we have integrated by parts and used that $\mathbf{U}$ is of compact support. Hence
\[
\langle\mathcal{L}_{_{^B\mathbf{K}_6}}\delta_{(U)}\!\left(X^2-Y^2\right)\mathbf{d}U\otimes\mathbf{d}U,\mathbf{U}\rangle=\int_{U=0}dV dX dY \,2\left(X^2-Y^2\right)\mathrm{U}^{UU}= 2\langle \delta_{(U)}\!\left(x^2-Y^2\right)\!\mathbf{d}U\otimes\mathbf{d}U,\mathbf{U}\rangle.
\]
Therefore
\begin{equation}
\langle\mathcal{L}_{_{^B\mathbf{K}_6}}{^B\mathbf{g}},\mathbf{U}\rangle=2\langle^B\mathbf{g},\mathbf{U}\rangle,\qquad\forall\, \mathbf{U}\in \mathcal{D}(\mathbb{R}^4,\otimes^2_0).
 \end{equation}
 
Next, consider the smooth the vector field ${^B\mathbf{K}}_7$ given by
\begin{equation}\label{conf_brinkmann_2}
^B\mathbf{K}_7= U^2\boldsymbol{\partial}_U+\frac{1}{2}\left(X^2+Y^2\right)\boldsymbol{\partial}_V +UX\boldsymbol{\partial}_X +UY\boldsymbol{\partial}_Y,
\end{equation}
As before, noting that $^B\mathbf{g}$ (\ref{pp_brinkmann}) can be written as in Eq.(\ref{pp_b_2}), and that
\begin{equation}\label{conf_mink}
\mathcal{L}_{_{^B\mathbf{K}_7}}\boldsymbol{\eta}=2 U \boldsymbol{\eta}\quad \Longrightarrow \quad
\langle \mathcal{L}_{_{^B\mathbf{K}_7}}{\boldsymbol{\eta}},\mathbf{U}\rangle=2 \langle U \boldsymbol{\eta}, \mathbf{U}\rangle,\qquad \forall\, \mathbf{U}\in \mathcal{D}(\mathbb{R}^4,\otimes^2_0),
\end{equation}
from (\ref{liederiv_a}) we see that we need to calculate
\[
\langle\mathcal{L}_{_{^B\mathbf{K}_7}}\delta_{(U)}\left(X^2-Y^2\right)\mathbf{d}U\otimes\mathbf{d}U,\mathbf{U}\rangle=-\left\langle \delta_{(U)}\!\left(X^2-Y^2\right)\mathbf{d}U\otimes\mathbf{d}U, \mathcal{L}_{_{^B\mathbf{K}_7}} \mathbf{U} +
\frac{1}{2}(\boldsymbol{g}^{-1}|\mathcal{L}_{_{{^B\mathbf{K}}_7}}\boldsymbol{g})\mathbf{U}\right\rangle.
\]
For $\boldsymbol{g}=\boldsymbol{\eta}$, we have $\frac{1}{2}(\boldsymbol{\eta}^{-1}|\mathcal{L}_{_{{^B\mathbf{K}}_7}}\boldsymbol{\eta})={^{\boldsymbol{\eta}}\nabla}\cdot{^B\mathbf{K}}_7=4U$ and
\[
\langle \delta_{(U)}\!\left(X^2-Y^2\right)\!\mathbf{d}U\otimes\mathbf{d}U, \mathcal{L}_{_{^B\mathbf{K}_7}} \mathbf{U} + \frac{1}{2}(\boldsymbol{\eta}^{-1}|\mathcal{L}_{_{{^B\mathbf{K}}_7}}\boldsymbol{\eta})\mathbf{U}\rangle=-\frac{1}{2}\int_{U=0}dV\,dX\,dY\,\partial_V\!\left[\left(X^2+Y^2\right){U}^{UU}\right]=0,
\]
where we have used $U\delta_{(U)}=0$ and the fact that $\mathbf{U}$ is of compact support. 
Therefore
\begin{equation}
\langle\mathcal{L}_{_{^B\mathbf{K}_7}}{^B\mathbf{g}},\mathbf{U}\rangle=2\langle U\,{^B\mathbf{g}},\mathbf{U}\rangle,\qquad \forall\, \mathbf{U}\in \mathcal{D}(\mathbb{R}^4,\otimes^2_0), 
 \end{equation}
i.e., $^B\mathbf{K}_7$ is a conformal Killing vector field of the distribution-valued tensor field $^B\mathbf{g}$ with conformal scalar $\psi=U$.
Finally, we note that the smooth vector fields ${^B\mathbf{K}}_6$ (\ref{conf_brinkmann_1}) and ${^B\mathbf{K}}_7$ (\ref{conf_brinkmann_2}) can be obtained formally from the vector fields ${^R\mathbf{K}}_6$ (\ref{conf_rosen_1}) and ${^R\mathbf{K}}_7$ (\ref{conf_rosen_2}), respectively, under the discontinuous transformation (\ref{trans}).

\section{Summary and Outlook}

By extending the notion of Lie derivative to distribution-valued tensor fields of order $m$, Lie derivatives with respect to vector fields of differentiability $C^k$, $k\geqslant m+1$, are defined. In terms of this extension, a (genuine linear) distributional approach is advocated for the study of the geometric symmetries of spacetimes of low differentiability.

As a specific application, we study the geometric symmetries of the impulsive plane wave spacetime with continuous metric ${^R\mathbf{g}}$ (\ref{pprosen_2}). Remarkably, the isometry algebra of ${^R\mathbf{g}}$ (\ref{pprosen_2}) contains non smooth vector fields of $C^1$ regularity, a result that was put forward in Ref.\cite{aichelburg96} and whose existence motivated in part this work. As was already mentioned, these non smooth Killing vector fields were obtained in that work as the formal pullback under (\ref{trans}) of the Killing vector fields (obtained by means of a regularization scheme) of the distributional form ${^B\mathbf{g}}$ (\ref{pp_brinkmann}). In this respect, our approach complements the one of Ref.\cite{aichelburg96}. Additionally, our analysis complete the study of the symmetries of the metric ${^R\mathbf{g}}$ (\ref{pprosen_2}) determining its conformal isometries. All the results are robust. The metric ${^R\mathbf{g}}$ (\ref{pprosen_2}) is within the maximal class of spacetime metrics allowing for a distributional curvature \cite{geroch,LeFloch:2007tv} and since this spacetime is Ricci-flat, we show that all the vector fields in its conformal algebra are special curvature collineations. We also determine the general form of the curvature collineations for this spacetime. Finally, we find the conformal Killing vector fields of the distributional form ${^B\mathbf{g}}$ (\ref{pp_brinkmann}), which turn out to be the formal pullback under (\ref{trans}) of the conformal Killing vector fields of the continuous metric ${^R\mathbf{g}}$ (\ref{pprosen_2}).  

Indeed, being based on the continuous extension of the Lie derivative (\ref{liederiv_b}), our approach is not restricted to a particular form of the metric tensor nor does it requires any regularization. Thus, as we have shown for a particular impulsive plane wave spacetime, symmetries generated by vector fields of regularity as low as $C^1$ in spacetimes of low differentiability can be rigorously considered \textit{via} the Lie derivative (\ref{liederiv_b}). We hope to return to this and other related issues in the near future.

\begin{acknowledgments}
J. C. is supported by ANID scholarship Nr. 21210008. We thank Alejandra Melfo for discussion and correspondence.
\end{acknowledgments}

\end{document}